\documentclass[11pt,letterpaper]{article}
\usepackage{jheppub}

 \usepackage{feynmp}

\usepackage{graphicx}
\usepackage{amsmath}
\usepackage{subfigure}
\usepackage{url}

\usepackage{multirow}
\usepackage{threeparttable}
\usepackage{paralist}

\newcommand{\MPl}{M_{\text{Pl}}}
\newcommand{\NL}{\text{NL}}

\newcommand{\bPhi}{{\boldsymbol \Phi}}

\newcommand{\bG}{{\boldsymbol G}}
\newcommand{\bU}{{\boldsymbol U}}
\newcommand{\bV}{{\boldsymbol V}}
\newcommand{\bH}{{\boldsymbol H}}
\newcommand{\bE}{{\boldsymbol E}}
\newcommand{\bC}{{\boldsymbol C}}
\newcommand{\bZ}{{\boldsymbol Z}}
\newcommand{\bK}{{\boldsymbol K}}
\newcommand{\bX}{{\boldsymbol X}}

\newcommand{\bQ}{{\boldsymbol Q}}
\newcommand{\bL}{{\boldsymbol L}}
\newcommand{\bS}{{\boldsymbol S}}

\newcommand{\bOmega}{{\boldsymbol \Omega}}
\newcommand{\bW}{{\boldsymbol W}}
\newcommand{\bR}{{\boldsymbol R}}

\newcommand{\bRtilde}{{\boldsymbol{\widetilde{R}}}}

\newcommand{\bWtilde}{{\boldsymbol{\widetilde{W}}}}
\newcommand{\bGtilde}{{\boldsymbol{\widetilde{G}}}}

\newcommand{\bSigma}{{\boldsymbol \Sigma}}
\newcommand{\bEpsilon}{{\boldsymbol{\mathcal{E}}}}
\newcommand{\FSW}{F_{\rm SW}}
\renewcommand{\P}{\mathcal{P}}
\newcommand{\bP}{{\boldsymbol{\mathcal{P}}}}

\newcommand{\vev}[1]{\left\langle #1 \right\rangle}
\newcommand{\nn}{\nonumber}
\newcommand{\gold}{\widetilde{G}_L}

\newcommand{\superbox}{\widetilde{\Box}}

\newcommand{\hc}{\text{h.c.}}

\renewcommand{\Re}{\mathop{\text{Re}}}

\newcommand{\Arg}{\mathop{\text{Arg}}}

\DeclareRobustCommand{\Sec}[1]{Sec.~\ref{#1}}

\DeclareRobustCommand{\App}[1]{App.~\ref{#1}}
\DeclareRobustCommand{\Tab}[1]{Table~\ref{#1}}

\DeclareRobustCommand{\Fig}[1]{Fig.~\ref{#1}}

\DeclareRobustCommand{\Eq}[1]{Eq.~(\ref{#1})}
\DeclareRobustCommand{\Eqs}[2]{Eqs.~(\ref{#1}) and (\ref{#2})}
\DeclareRobustCommand{\Ref}[1]{Ref.~\cite{#1}}
\DeclareRobustCommand{\Refs}[1]{Refs.~\cite{#1}}

\newcommand{\be}{\begin{equation}}
\newcommand{\ee}{\end{equation}}

\newcommand{\alphadot}{{\dot{\alpha}}}
\newcommand{\betadot}{{\dot{\beta}}}
\newcommand{\gammadot}{{\dot{\gamma}}}

\newcommand{\X}{{\bar{X}}}
\renewcommand{\i}{{\bar{\imath}}}
\renewcommand{\j}{{\bar{\jmath}}}

\renewcommand{\l}{{\bar{l}}}
\newcommand{\D}{{\mathcal{D}}}
\newcommand{\sigmabar}{\overline{\sigma}}

\title{Anomaly Mediation from Unbroken Supergravity}

\author[a,b]{Francesco D'Eramo,}
\author[c]{Jesse Thaler,}
\author[c]{and Zachary Thomas}

\affiliation[a]{Department of Physics, University of California, Berkeley, CA 94720, USA}
\affiliation[b]{Theoretical Physics Group, Lawrence Berkeley National Laboratory, Berkeley, CA 94720,
USA}
\affiliation[c]{Center for Theoretical Physics, Massachusetts Institute of Technology, Cambridge, MA 02139, USA}

\emailAdd{fraderamo@berkeley.edu}
\emailAdd{jthaler@mit.edu}
\emailAdd{ztt@mit.edu}

\abstract{When supergravity (SUGRA) is spontaneously broken, it is well known that anomaly mediation generates sparticle soft masses proportional to the gravitino mass.  Recently, we showed that one-loop anomaly-mediated gaugino masses should be associated with unbroken supersymmetry (SUSY).  This counterintuitive result arises because the underlying symmetry structure of (broken) SUGRA in flat space is in fact (unbroken) SUSY in anti-de Sitter (AdS) space.  When quantum corrections are regulated in a way that preserves SUGRA, the underlying AdS curvature (proportional to the gravitino mass) necessarily appears in the regulated action, yielding soft masses without corresponding goldstino couplings.    In this paper, we extend our analysis of anomaly mediation to sfermion soft masses.  Already at tree-level we encounter a number of surprises, including the fact that zero soft masses correspond to broken (AdS) SUSY.   At one-loop, we explain how anomaly mediation appears when regulating SUGRA in a way that preserves super-Weyl invariance.  We find that recent claims in the literature about the non-existence of anomaly mediation were based on a Wilsonian effective action with residual gauge dependence, and the gauge-invariant 1PI effective action contains the expected anomaly-mediated spectrum.  Finally, we calculate the sfermion spectrum to all orders, and use supertrace relations to derive the familiar two-loop soft masses from minimal anomaly mediation, as well as unfamiliar tree-level and one-loop goldstino couplings consistent with renormalization group invariance.}


\begin{document}

\hfill MIT-CTP 4459 

\hfill UCB-PTH-13/05

\maketitle

\section{Introduction}
If supersymmetry (SUSY) is realized in nature, then it must be spontaneously broken.  Spontaneously broken SUSY yields a positive contribution to the cosmological constant, so in order to achieve the nearly zero cosmological constant we see today, the underlying symmetry structure of our universe must be SUSY in anti-de Sitter (AdS) space.   In the context of supergravity (SUGRA), the inverse AdS radius $\lambda_{\rm AdS}^{-1}$ is equal to the gravitino mass $m_{3/2}$.  Thus, because of the underlying AdS SUSY algebra, there will be effects on the supersymmetric standard model (SSM) proportional to $m_{3/2}$.  These would appear as ``SUSY-breaking'' effects from the point of view of the flat space SUSY algebra, but are actually SUSY-preserving effects when viewed from AdS$_4$ space.

Famously, anomaly mediation \cite{Randall:1998uk,Giudice:1998xp} yields gaugino masses proportional to $m_{3/2}$.  As we recently showed in \Ref{D'Eramo:2012qd}, these gaugino masses do not break AdS SUSY, and are in fact necessary for conservation of the AdS supercurrent.  We called this phenomenon ``gravitino mediation'' to separate this $m_{3/2}$ effect from other anomaly-mediated effects which have nothing to do with the AdS SUSY algebra.\footnote{These other effects were dubbed ``K\"ahler mediation'' since they arise from linear couplings of SUSY breaking to visible sector fields in the K\"ahler potential.  Full anomaly mediation is simply the sum of K\"ahler mediation and gravitino mediation.  See \Ref{D'Eramo:2012qd} for details.  There is also a (usually subleading) anomaly-mediated effect noted in \Ref{Conlon:2010qy} if there are direct couplings of SUSY breaking to the gauginos at tree-level.}   Throughout this paper, we will use the more familiar (but less accurate) name ``anomaly mediation'' to refer to all effects proportional to $m_{3/2}$ (i.e.~gravitino mediation; see \Refs{Chacko:1999am,Bagger:1999rd,Bagger:2000dh,Dine:2007me,Gripaios:2008rg,Jung:2009dg,Conlon:2010qy,Sanford:2010hc} for additional theoretical perspectives).  Unlike usual SUSY-breaking effects, anomaly mediation generates gaugino masses without accompanying goldstino couplings, further emphasizing that this is a SUSY-preserving effect.

The goal of this paper is twofold.  First, we wish to extend the analysis of \Ref{D'Eramo:2012qd} to the case of sfermions.  It is well known that anomaly mediation yields two-loop scalar mass-squareds proportional to $m_{3/2}^2$, but we will show that from the point of view of AdS$_4$ space, anomaly mediation already yields scalar masses at tree level.  Following the strategy of \Ref{D'Eramo:2012qd}, we will use goldstino couplings as a guide to determine which effects preserve AdS SUSY, allowing us to distinguish between SUSY-preserving effects that are genuinely proportional to $m_{3/2}$ versus SUSY-breaking effects that are only proportional to $m_{3/2}$ because of the need to fine tune the cosmological constant to zero.  Second, we wish to counter recent claims by de Alwis that anomaly mediation does not exist \cite{deAlwis:2008aq,deAlwis:2012gr}.  In contrast, we will use the same logical starting point as de Alwis (which is based on the analysis of Kaplunovsky and Louis \cite{Kaplunovsky:1994fg}) but come to the conclusion that anomaly mediation not only exists, but is necessary for the preservation of AdS SUSY.  

Along the way, we will encounter a number of surprises, all ultimately having to do with the structure of AdS SUSY:
\begin{itemize}
\item \textbf{Tree-Level Tachyons and Sequestering}.  Already at tree-level in AdS space, the components of a chiral multiplet get SUSY mass splittings proportional to $m_{3/2}$.  For example, if the fermionic component is massless, then its scalar partner has a negative mass-squared $-2m_{3/2}^2$, satisfying the Breitenlohner-Freedman bound \cite{Breitenlohner:1982jf}.\footnote{A fermion with mass $\pm \frac{1}{2}m_{3/2}$ will have one scalar partner with mass-squared $-\frac{9}{4} m_{3/2}^2$, exactly saturating the bound.}  In order to have a stable theory after AdS SUSY is lifted to flat space via SUSY breaking, this negative mass-squared must also be lifted.  Since such a lifting must break AdS SUSY, this requires irreducible couplings between the SUSY-breaking sector (``hidden sector'') and the SSM (``visible sector''), even in theories where the hidden and visible sectors are sequestered \cite{Randall:1998uk}.  For a chiral multiplet with components $\{\phi, \chi, F\}$ there is necessarily a coupling to the goldstino $\gold$ when the sfermion soft mass is zero in flat space:
\be
\label{eq:irreduciblegoldstinoIntro}
\mathcal{L} \supset \frac{2 m_{3/2}^2}{F_{\rm eff}} \gold \chi \phi^*,
\ee
where $F_{\rm eff}$ is the scale of SUSY breaking.  Intriguingly, this coupling is renormalization-group invariant, and effectively defines what it means to sequester the hidden and visible sectors.\footnote{In \Ref{D'Eramo:2012qd}, we (erroneously) advocated that the absence of goldstino couplings could be used as a physical definition of sequestering.  Because of this tree-level tachyon subtlety, though, this goldstino coupling is needed to have a stable theory.}

\item \textbf{Giudice-Masiero in AdS Space}.  In flat space, the harmonic part of the K\"ahler potential (i.e.~the chiral plus anti-chiral part) is unphysical.  This is not the case in AdS space, and the Giudice-Masiero mechanism~\cite{Giudice:1988yz} is a way to generate $\mu$ and $B_\mu$ terms via $\bK \supset \bH_u \bH_d + \text{h.c.}$  While the generated $\mu$ term preserves AdS SUSY, the $B_\mu$ term actually breaks AdS SUSY, since it secretly involves direct couplings between Higgs multiplets and the goldstino.  When written in a more natural basis, it becomes clear that Giudice-Masiero arises from a combination of a SUSY-preserving and SUSY-breaking effect.

\item \textbf{Anomaly Mediation and Super-Weyl Invariance}.  As emphasized in \Ref{Dine:2007me}, anomaly mediation is not due to any anomaly of SUSY itself,\footnote{Of course, the name ``anomaly mediation'' is still justified since it generates effects proportional to beta function coefficients.} but is rather due to the need to add local counterterms to preserve SUSY of the 1PI effective action.  A related story presented in \Ref{Gripaios:2008rg} is that bulk counterterms are needed to counteract otherwise SUSY-breaking effects due to the boundary of AdS$_4$.  Here, we will follow the logic of de Alwis \cite{deAlwis:2008aq,deAlwis:2012gr} (based on the analysis of Kaplunovsky and Louis \cite{Kaplunovsky:1994fg}) to show how anomaly mediation arises from preserving super-Weyl invariance of a UV-regulated SUGRA theory.  While de Alwis (erroneously) concluded that anomaly mediation cannot exist in such a situation, we find that there is residual gauge dependence in de Alwis' calculation  (and a similar issue implicit in Kaplunovsky and Louis).  In the langauge of the Weyl compensator, anomaly mediation depends not just on the $F_C$ component of the compensator (which can be gauge-fixed to zero), but on the super-Weyl-invariant combination
\be
F_{\rm SW} \equiv F_C - \frac{1}{3}M^*,
\ee
where $M$ is the scalar auxiliary field.   Accounting for the fact that $\vev{F_{\rm SW}}$ depends on $m_{3/2}$, we reproduce the familiar anomaly-mediated spectrum.  
 \item \textbf{Supertraces Resolve Spectrum Ambiguities}.  We will use an ansatz for the SUGRA-invariant 1PI effective action to extract sfermion soft masses and goldstino couplings.  Because there are many such ans\"atze consistent with SUGRA, there is an ambiguity in the resulting sfermion spectrum.  For example, there are three terms that show up at $\mathcal{O}(m_{3/2}^2)$ in the 1PI effective action:
\begin{align}
\mathcal{L}_\text{soft mass} & = - \mathcal{C}_s \phi^* \phi - \mathcal{C}_a F^* \Box^{-1} F + i \mathcal{C}_f \chi^\dagger \sigmabar^\mu \D_\mu \Box^{-1} \chi,
\end{align}
where $\Box$ is the d'Alembertian appropriate to curved space.  The first term is the familiar sfermion soft mass-squared term, but the two non-local terms necessarily appear as $m^2/p^2$ corrections to the self-energies.  We will find that while the coefficients $\mathcal{C}_i$ are indeed ambiguous (since they depend the precise form of the ansatz), the supertrace
\begin{align}
\mathcal{S} & = \mathcal{C}_s + \mathcal{C}_a - 2 \mathcal{C}_f
\end{align}
is unambiguous and gives a useful measure of the ``soft mass-squared'' for a sfermion (see \Ref{ArkaniHamed:1998kj} for a related story).  Not surprisingly, a similar supertrace is needed to define unambiguous ``goldstino couplings''.
\item  \textbf{SUSY-Breaking in the SUGRA Multiplet}.  The key confusion surrounding anomaly mediation is that there are two different order parameters in SUGRA---one which sets the underlying AdS curvature and one which accounts for SUSY breaking---which are only related to each other after tuning the cosmological constant to zero.  In particular, a non-vanishing vacuum expectation value (vev) for $M^*$ (containing the term $-3 m_{3/2}$ in SUGRA frame) does not break SUSY.  Instead, the SUSY-breaking order parameter in SUGRA comes from the $F$-component of the chiral curvature superfield $\bR$:
\begin{align}
F_R & \equiv \frac{1}{12} \mathcal{R} - m_{3/2}^2.
\end{align}
After using the Einstein equation, $F_R$ vanishes for unbroken SUSY in AdS, but takes on the value $-m_{3/2}^2$ once the cosmological constant has been tuned to zero.  Thus in flat space, we will find both SUSY-breaking and SUSY-preserving effects proportional to $m_{3/2}^2$, and we will have to tease these two effects apart by carefully considering AdS SUSY.  We will also find corresponding goldstino couplings proportional to $F_R$, arising from terms in the SUGRA multiplet proportional to the gravitino equations of motion.  
\item \textbf{Two-Loop Soft Masses and One-Loop Goldstino Couplings}. Using an ansatz for the all-orders SUGRA-invariant 1PI effective action, we will recover the familiar two-loop soft masses from anomaly mediation.  But in addition, we will find one-loop goldstino couplings proportional to anomalous dimensions (on top of the tree-level goldstino coupling from \Eq{eq:irreduciblegoldstinoIntro}).  As a cross check of our calculation, both the two-loop soft mass and the one-loop goldstino coupling are renormalization-group (RG) invariant quantities, as expected from the general analysis of \Refs{Jack:1997eh,Jack:1999aj,Pomarol:1999ie,ArkaniHamed:1998kj}.  The complete sfermion spectrum is summarized in \Tab{tab:summary}. 

\end{itemize}

\renewcommand{\arraystretch}{1.5}
\begin{table}
\begin{center}
\begin{tabular}{|c|c||ccc|}
\hline
& & Tree-Level & One-Loop & Two-Loop \\
\hline \hline
 \multirow{2}{*}{\parbox{2.8cm}{\centering \textbf{SUSY AdS$_4$} \\ ($\mathcal{R} = 12 m^2_{3/2}$)}} & Soft Mass-Squared & $-2 m^2_{3/2}$ & $\gamma m^2_{3/2}$ & $- \frac{1}{4} \dot{\gamma} m_{3/2}^2$\\
 & Goldstino Coupling & --- & --- & --- \\
 \hline
 \hline
 \multirow{2}{*}{\parbox{2.8cm}{\centering \textbf{Curved Space} \\ (broken SUSY)}} & Soft Mass-Squared & $-\frac{1}{6}\mathcal{R}$ & $\frac{1}{12}\gamma \mathcal{R}$ & $- \frac{1}{4} \dot{\gamma} m_{3/2}^2$\\
 & Goldstino Coupling & $-2(m^2_{3/2} - \frac{1}{12}\mathcal{R})$ & $\gamma(m^2_{3/2}-\frac{1}{12}\mathcal{R})$ & --- \\
 \hline
 \hline
  \multirow{2}{*}{\parbox{2.8cm}{\centering \textbf{Flat Space} \\ (broken SUSY)}} & Soft Mass-Squared & --- & --- & $- \frac{1}{4} \dot{\gamma} m_{3/2}^2$\\
 & Goldstino Coupling & $-2 m^2_{3/2}$ & $\gamma m^2_{3/2}$ & --- \\
 \hline
 \hline
\end{tabular}
\end{center}
\caption{Sfermion soft masses and goldstino couplings from minimal anomaly mediation (i.e.~``gravitino mediation'' in the language of \Ref{D'Eramo:2012qd}, so $\vev{K_i} = 0$).  Here, $\gamma$ is the anomalous dimension of the chiral multiplet and $\dot{\gamma} \equiv d \gamma / d \log \mu$.  Starting with unbroken SUSY in AdS$_4$ with Ricci curvature $\mathcal{R} =  12 \lambda_{\rm AdS}^{-2} = 12 m^2_{3/2}$, we show how the spectrum evolves as SUSY breaking is tuned to achieve flat space with $\mathcal{R} \to 0$.  In this table, ``soft mass-squared'' and ``goldstino coupling'' refer to the supertraces in \Eqs{eq:supertrace}{eq:supertracegoldstinoterms}, and the loop level refers to the order at which the effect starts.  Minimal anomaly mediation also yields $A$-terms and $B$-terms, which are described in \Sec{subsec:finalanswer}.  This table only includes the contributions from bulk terms and not from one- and two-loop boundary terms (analogous to \Ref{Gripaios:2008rg}) necessary to preserve the SUSY algebra in AdS$_4$; these boundary terms are irrelevant in flat space.}
\label{tab:summary}
\end{table}

The remainder of this paper is organized as follows.  In \Sec{sec:treelevel}, we review the structure of SUGRA at tree-level, and show how the underlying AdS algebra gives rise to SUSY-preserving mass splittings between fermions and sfermions.  In \Sec{sec:oneloop}, we discuss super-Weyl invariance in UV-regulated SUGRA theories at one loop, and show how anomaly mediation arises as a super-Weyl-preserving and SUSY-preserving effect.  In \Sec{sec:twoloop}, we discuss anomaly mediation for sfermions up to two-loop order, completing the analysis of goldstino couplings that was initiated in \Ref{D'Eramo:2012qd}.  We conclude in \Sec{sec:discuss}.

\section{Invitation:  Anomaly Mediation at Tree Level}
\label{sec:treelevel}

\begin{figure}
\centering
\includegraphics[width=4.0in]{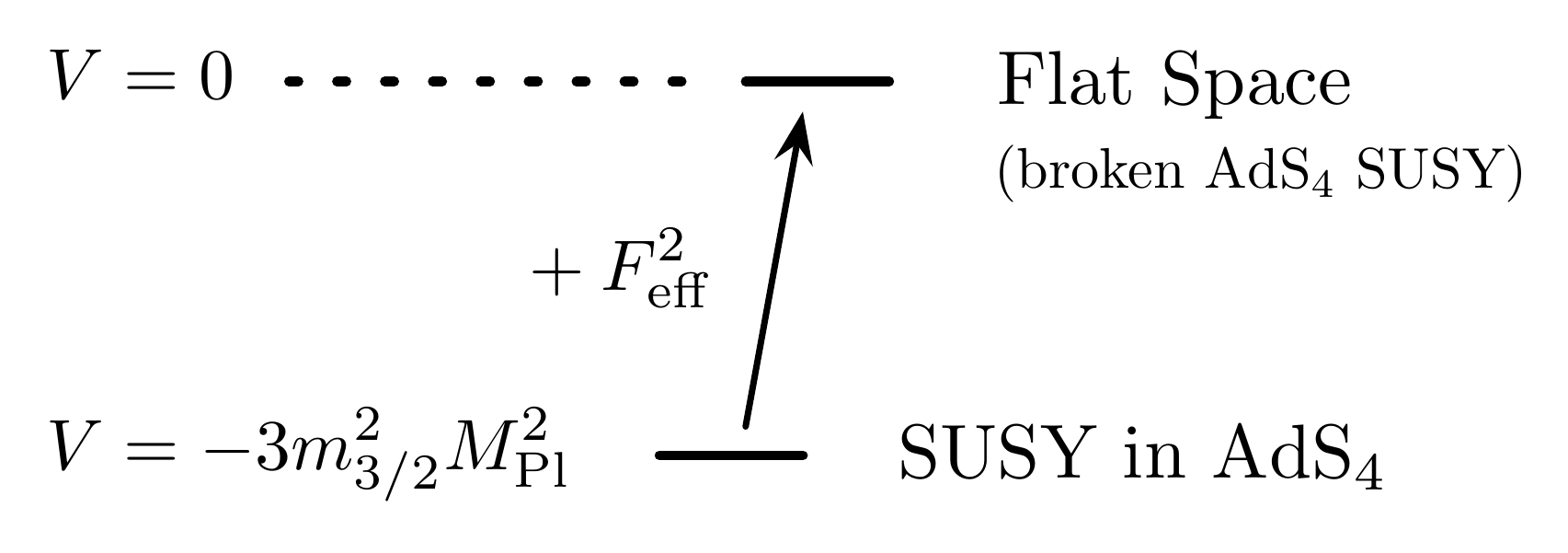}
\caption{Fine-tuning of the cosmological constant, adapted from \Ref{Bertolini:2013via}.  Starting with the underlying AdS radius $\lambda_{\rm AdS}^{-1} = m_{3/2}$, SUSY-breaking effects lead to flat space with broken (AdS) SUSY.}
\label{fig:tuning}
\end{figure}

It is well known that rigid AdS SUSY requires mass splittings between particles and sparticles \cite{Breitenlohner:1982jf,Nicolai:1984hb}.  Less well known is that those mass splittings have an impact on the phenomenology of SUGRA, even if the geometry (after SUSY breaking) is that of flat space.  In particular, the couplings of the goldstino (eaten to form the longitudinal components of the gravitino) can be used to track which effects break SUSY and which effects preserve SUSY.   Crucially, these couplings depends on $m_{3/2}$, which in turn depends on the underlying AdS radius $\lambda_{\rm AdS}^{-1} = m_{3/2}$ \emph{prior} to SUSY breaking.  The fine-tuning of the cosmological constant to achieve flat space is summarized in \Fig{fig:tuning}.

Considering only chiral multiplets, we can write the fermion and sfermion masses and sfermion-fermion-goldstino couplings as
\be
\mathcal{L} \supset - m^2_{\i j} \phi^{* \i} \phi^j - \frac{1}{2} B_{ij} \phi^i \phi^j - \frac{1}{2} M_{ij} \chi^i \chi^j + \frac{a_{\i j}}{F_{\rm eff }} \phi^{* \i} \chi^j \gold + \frac{b_{ij}}{F_{\rm eff }} \phi^i \chi^j \gold + \text{h.c.}, \label{eq:massesandcouplings}
\ee
where $\phi_i$ is a sfermion, $\psi_i$ is its fermion partner, $\gold$ is the goldstino, and $F_{\rm eff}$ is the scale of SUSY breaking.    Assuming the flat space SUSY algebra, one can show that
\be
a^{\rm flat}_{\i j} = m^2_{\i j} - M_{\i}{}^k M_{k j}, \qquad b^{\rm flat}_{ij} = B_{ij},
\ee
which emphasizes that goldstino couplings arise when sfermions and fermions have non-zero mass splittings (i.e.~when flat space SUSY is broken).  In AdS space at tree-level, however, we will show that
\be
a^{\rm AdS}_{\i j} = m^2_{\i j} - M_{\i}{}^k M_{k j} + 2 m_{3/2}^2 \delta_{\i j}, \qquad b^{\rm AdS}_{ij} = B_{ij}  + m_{3/2} M_{ij},
\label{eq:goldstinocouplings}
\ee
which shows that one can have $m_{3/2}$-dependent mass splittings between multiplets without corresponding goldstino couplings (i.e.~without breaking AdS SUSY).  

In this section, we give two different derivations of \Eq{eq:goldstinocouplings}, with a third derivation using the conformal compensator given in \App{app:compensator}.  We then discuss the phenomenological implications of these goldstino couplings for sequestering, Giudice-Masiero terms, and regulator fields.  Though the goldstino is eaten by the gravitino in SUGRA, the couplings of the goldstino are still physically relevant.  Indeed, in the goldstino equivalence theorem regime with energies $E \gg m_{3/2}$, the interactions of the longitudinal components of the gravitino are captured by the goldstino couplings in \Eq{eq:goldstinocouplings} (plus modifications to those goldstino couplings that appear at higher-loop order).  
\subsection{Derivation from the SUGRA Lagrangian}

\label{subsec:SUGRAlagrangian}

The first way to derive \Eq{eq:goldstinocouplings} is to consider the SUGRA lagrangian directly.  The scalar potential for SUGRA  is  \cite{wess1992supersymmetry}
\be
V = e^G (G^k G_k - 3 ),
\ee
where the  K\"ahler-invariant potential $G$ is given by\footnote{The K\"ahler anomaly \cite{Bagger:1999rd,Bagger:2000dh} implies a physical difference between the K\"ahler potential and the superpotential, but it does not enter at tree level.}
\be
\label{eq:Gdef}
G \equiv K + \log W + \log W^\dagger.
\ee
Throughout the text, we use the conventions of \Ref{wess1992supersymmetry}.  Here, subscripts represent derivatives with respect to scalar fields ($G_k = \partial G / \partial \phi^k$), and indices are raised and lowered with the K\"ahler metric $G_{i \j}$ and its inverse.  The gravitino mass is given by
\be
m_{3/2} = \vev{e^{G/2}},
\ee
and the quadratic fermion interactions in SUGRA are
\be
\mathcal{L} \supset - i G_{i \j} \chi^{\dagger \j} \sigmabar^\mu \mathcal{D}_\mu \chi^i - \frac{1}{2} e^{G/2} (\nabla_i G_j + G_i G_j) \chi^i \chi^j + \rm h.c. \label{eq:fermionmassesSUGRAalt}
\ee
where $\D_\mu$ and $\nabla_i$ are the K\"ahler-covariant derivatives with respect to spacetime and scalar fields, respectively.  

If SUGRA is unbroken ($\vev{G_i} = 0$), then we have a negative cosmological constant ($\vev{V} = -3 m_{3/2}^2 \MPl^2$), so the spacetime background is AdS, with curvature $\lambda_{\rm AdS}^{-1} = m_{3/2}$.  The fermion mass matrix is
\be
M_{i j} = m_{3/2} \vev{\nabla_i G_j} \qquad (\text{unbroken SUGRA}),
\ee
and at the extremum of the potential ($\vev{V_i} = 0$), the scalar mass-squared and holomorphic mass can be expressed in terms of $M_{ij}$ as
\begin{align}
m^2_{i \j} & =  M_{i k} M^k{}_\j - 2 m_{3/2}^2 \delta_{i \j}, \label{eq:unbrokenSUSYspectrum} \\
B_{i j} & = - m_{3/2} M_{i j}  \qquad (\text{unbroken SUGRA}).
\end{align}
Note that inserting these mass values into \Eq{eq:goldstinocouplings} yields no goldstino couplings,  as is to be expected since there is no goldstino when SUGRA is unbroken.

If SUGRA is broken, then there are a few important effects.  Defining the SUSY-breaking scale as 
\be
F_{\rm eff} \equiv \sqrt{e^G G^k G_k}, 
\ee
we find the the cosmological constant is modified to be
\begin{align}
\vev{V} & = F_{\rm eff}^2 - 3 m_{3/2}^2 \MPl^2,  \label{eq:cosmologicalconstant}
\end{align}
where we have restored factors of the Planck constant $\MPl$.  As shown in \Fig{fig:tuning}, it is possible to fine-tune $V=0$ by choosing
\be
F_{\rm eff} = \sqrt{3} m_{3/2} M_{\rm Pl}.   
\ee
In addition, SUSY breaking gives rise to a goldstino, which (assuming no $D$-terms for the gauge multiplets for simplicity) points in the direction
\be
\gold = - \frac{1}{\sqrt{3}} G^i \chi_i.
\ee
 The fermion and sfermion mass matrices are generically deformed due to the presence of SUSY breaking, and their form is well-known for $\vev{V} = 0$ and $\vev{V_i} = 0$ \cite{wess1992supersymmetry}:\footnote{There is a typo in \Ref{wess1992supersymmetry} which omits the first term in \Eq{eq:BtermsSUGRA}.}
 \begin{align}
M_{i j} & = m_{3/2} \vev{\nabla_i G_j + G_i G_j}, \\
m^2_{i \j} & = m_{3/2}^2 \vev{ \nabla_i G_k \nabla_\j G^k - R_{i \j k \l} G^k G^\l + G_{i \j} }, \label{eq:scalarmassesSUGRA} \\
m^2_{i j} & = m_{3/2}^2 \vev{ G^k \nabla_i \nabla_j G_k + 2 \nabla_i G_j }, \label{eq:BtermsSUGRA}
\end{align}
where $R_{i \j k \l}$ is the K\"ahler curvature tensor.\footnote{Here, and throughout the text, we do not choose any gauge fixing for the gravitino, so there is also quadratic mixing between the goldstino and the gravitino.  See \Eq{eq:effectiveGoldstinolagrangian} below.}

The Yukawa couplings can similarly be extracted from \Eq{eq:fermionmassesSUGRAalt}: 
\begin{align}
\mathcal{L} & \supset - \frac{1}{2} m_{3/2} \vev{- R_{i \j k \l} G^\l + G_{i \j} G_k + G_i G_{k \j} } \chi^i \chi^k \phi^{* \j} \\
& \quad \, - \frac{1}{2} m_{3/2} \vev{\nabla_i \nabla_j G_k + G_i \nabla_j G_k + G_k \nabla_i G_j + G_j \nabla_k G_i + G_i G_j G_k} \chi^i \chi^k \phi^j.  \label{eq:SUSYyukawa}
\end{align}
One can read off the couplings of the goldstino to visible-sector fields after picking out the goldstino direction: 
\begin{align}
a_{\i j} &= m_{3/2}^2 \vev{- R_{i \j k \l} G^k G^\l + 3 G_{i \j} },\\
b_{ij} & = m_{3/2}^2 \vev{G^k \nabla_i \nabla_j G_k + 3 \nabla_i G_j },
\end{align}
recalling that $\vev{G_i}$ is negligible for visible-sector fields.  This then yields the goldstino couplings anticipated in \Eq{eq:goldstinocouplings} (at least for the case of $\vev{V} = 0$).

Thus, despite the fact that SUGRA is broken and the cosmological constant is lifted to yield $\vev{V} = 0$, the goldstino couplings retain information about the structure of the underlying AdS SUSY, and not the structure of flat space SUSY.

\subsection{Derivation from Supercurrent Conservation}
\label{subsec:supercurrent}

An alternative derivation of \Eq{eq:goldstinocouplings} uses conservation of the AdS supercurrent.  The supercurrent is the Noether current of (rigid) SUSY transformations, and in SUGRA, the linear couplings of the gravitino $\psi_\mu$ to matter are determined by the supercurrent alone:
\begin{align}
\label{eq:gravitinoLag}
\mathcal{L} & = \epsilon^{\mu \nu \rho \tau} \psi^\dagger_\mu \sigmabar_\nu \D_\rho \psi_\tau - m_{3/2} \psi^\dagger_\mu \sigmabar^{\mu \nu} \psi^\dagger_\nu + \textrm{h.c.} - \frac{1}{2 \MPl} \psi^{\dagger}_\mu j^{\dagger \mu} + \rm h.c. 
\end{align}
Appropriate manipulation of the gravitino equation of motion (and the Einstein equation, given \Eq{eq:cosmologicalconstant}) yields the relation
\begin{align}
0 & =  \left( \D_\mu j^{\dagger \mu} + \frac{1}{2} i m_{3/2} \sigmabar^\mu j_\mu \right) -  i  \frac{F_{\rm eff}^2}{\MPl} \sigmabar^\mu \psi_\mu. \label{eq:SUGRAsupercurrentconservation}
\end{align}
This relation can be most naturally interpreted in the rigid limit ($\MPl \rightarrow \infty$, $m_{3/2}$ and $F_{\rm eff}$ fixed), in which the last term vanishes and the spacetime background is AdS (with $\lambda_{\rm AdS}^{-1} = m_{3/2}$).  In the rigid limit, we see clearly that conservation of the supercurrent is different in flat space versus AdS space.  In flat space, the fermionic SUSY transformation parameter $\epsilon$ satisfies the criteria $\partial_\mu \epsilon = 0$, whereas in AdS space
\be
\D_\mu \epsilon = - \frac{i}{2} m_{3/2} \sigma_\mu \epsilon^\dagger, \label{eq:epsilonconstraint}
\ee
where $\D_\mu$ is the (gravity) covariant derivative \cite{Adams:2011vw,Festuccia:2011ws}.  Among other things, this implies that the goldstino in rigid AdS space has a mass of $2m_{3/2}$ \cite{Cheung:2010mc,Cheung:2011jq}.  It also implies that the condition for conservation of the supercurrent is not $\partial_\mu j^\mu = 0$ but rather the rigid limit of \Eq{eq:SUGRAsupercurrentconservation}, as Noether's theorem requires $\D_\mu (j^\mu \epsilon + j^{\dagger \mu } \epsilon^\dagger) = 0$. 

When SUSY is broken, the supercurrent contains the goldstino
\be
\label{eq:supercurrentWithGold}
 j^{\dagger \mu} = \sqrt{2} F_{\rm eff} i \sigmabar^\mu \gold + \widetilde{j}^{\dagger \mu},
\ee
where $\widetilde{j}_{\mu}$ is the remaining ``matter'' part of the supercurrent.  \Eq{eq:SUGRAsupercurrentconservation} can then be interpreted as the goldstino equation of motion arising from the lagrangian 
\begin{align}
\mathcal{L} & = - i \gold^\dagger \sigmabar^\mu \nabla_\mu \gold - \frac{1}{2} (2 m_{3/2}) \gold \gold + \hc + \frac{i}{\sqrt{2}} \frac{F_{\rm eff}}{\MPl} \gold^\dagger \sigmabar^\mu \psi_\mu + \hc \nonumber \\
& \quad \, - \frac{1}{\sqrt{2} F_{\rm eff}} \left( \D_\mu \widetilde{j}^\mu - \frac{1}{2} i m_{3/2} \widetilde{j}^{\dagger \mu} \sigmabar_\mu \right) \gold + \hc, \label{eq:effectiveGoldstinolagrangian}
\end{align}
where the last term is necessary for conservation of the AdS supercurrent.

In both flat space and AdS space, the supercurrent for chiral multiplets contains\footnote{This assumes that the SUGRA action only contains a K\"ahler potential and a superpotential without additional higher-derivative interactions.  The supercurrent is modified when loop effects are taken into account, giving rise to new effects detailed in \Sec{sec:twoloop}.}
\be
j^\mu \supset \sqrt{2} g_{i \j} \partial_\nu \phi^{* \j} \chi^i \sigma^\mu \sigmabar^\nu.  \label{eq:supercurrent}
\ee
The other term proportional to $\chi^{\dagger \i} D_\i W^* \chi^{\dagger \i} \sigmabar^\mu$ is irrelevant for our discussions since it vanishes  on the goldstino equation of motion.  Using the equations of motion for the matter fields and the goldstino equation of motion, we find that \Eq{eq:effectiveGoldstinolagrangian} contains the goldstino couplings
\begin{align}
a_{i \j} & = m^2_{i \j} - M_{i k} M^k{}_\j + 2 m_{3/2}^2 \delta_{i \j}, \\
b_{ij} & =  B_{i j} + m_{3/2} M_{i j} ,
\end{align}
as expected from \Eq{eq:goldstinocouplings}.  Note that the terms proportional to $m_{3/2}$ arise from the additional goldstino mass and $\frac{1}{F_{\rm eff}} i m_{3/2} \widetilde{j}^{\dagger \mu} \sigmabar_\mu \gold$ terms necessary for AdS supercurrent conservation.

\subsection{Tachyonic Scalars and Sequestering}

\begin{figure}
\centering
\includegraphics[width=3.0in]{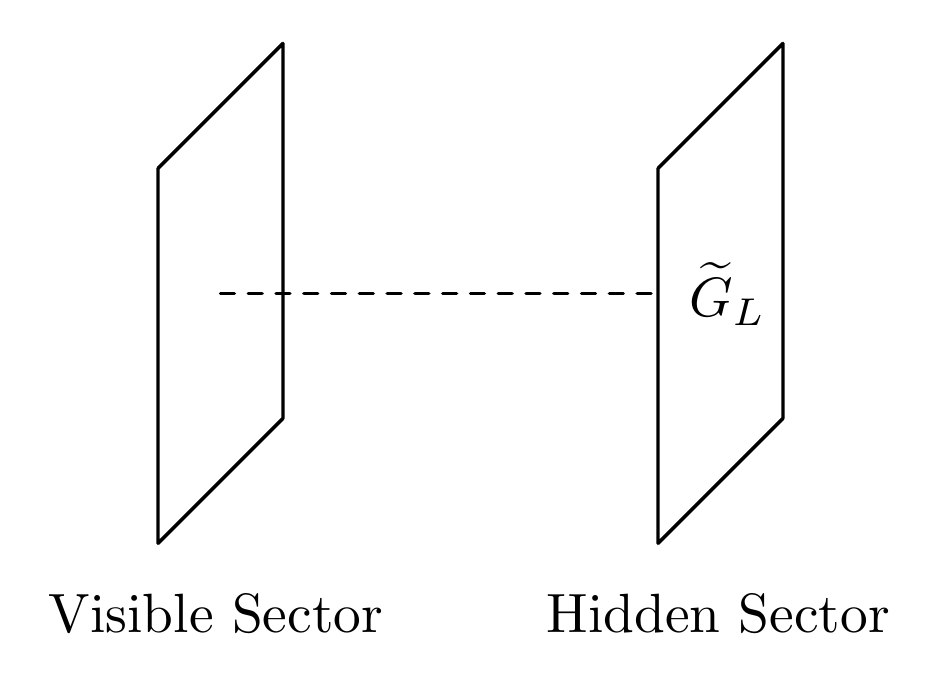}
\caption{An extra-dimensional realization of the sequestered limit, where SUSY is broken only in a hidden sector.  Naively, the goldstino is localized in the hidden sector and would not couple to visible sector fields.  But due to mixing with the gravitino, there are irreducible couplings between the goldstino and chiral multiplets in the visible sector in order to have a stable tree-level theory in flat space after SUSY breaking.}
\label{fig:sequestered}
\end{figure}

The fermions in the standard model are massless (prior to electroweak symmetry breaking), so in the absence of AdS SUSY breaking, the sfermions would be tachyonic, with a common mass-squared $-2 m_{3/2}^2$ (see \Eq{eq:unbrokenSUSYspectrum}).  In order to have a (meta)stable vacuum after SUSY breaking, these tachyonic masses must be lifted, but from the $a_{i \j}$ term in \Eq{eq:goldstinocouplings}, this implies an irreducible coupling between the goldstino and the matter fields.

This result is rather surprising from the point of view of strictly sequestered theories \cite{Randall:1998uk}, where anomaly mediation is the only source of soft masses.  As shown in \Fig{fig:sequestered}, one way to achieve the sequestered limit is to have the visible sector (i.e.~the SSM) and the hidden sector~(i.e.~SUSY-breaking dynamics) live in different parts of an extra-dimensional space with no light degrees of freedom connecting the two apart from gravity.  This implies a special sequestered form of the effective four-dimensional K\"ahler potential and superpotential:
\be
\label{eq:factorized_form}
-3 e^{- K/3} = \Omega_{\rm vis} +\Omega_{\rm hid}, \qquad W = W_{\rm vis} + W_{\rm hid}.
\ee
Naively, one would think that the goldstino from SUSY-breaking must be localized in the hidden sector (assuming the SSM itself does not break SUSY \cite{Izawa:2011hi,Bertolini:2011tw}), and therefore decoupled from the visible sector.  But \Eq{eq:goldstinocouplings} shows that there are direct connections between the visible and hidden sectors necessary for stability of the theory.  In particular, there is an irreducible coupling to the goldstino when the sfermion soft mass is zero in flat space:
\be
\label{eq:irreduciblegoldstino}
\mathcal{L} \supset \frac{2 m_{3/2}^2}{F_{\rm eff}} \gold \chi \phi^*.
\ee

There are two potential ways to interpret this result.  One interpretation is to conclude that sequestering corresponds to a fine-tuned limit.   After all, in the sequestered limit at tree-level, one has the underlying $-2 m_{3/2}^2$ AdS tachyonic mass balanced against the $+2 m_{3/2}^2$ SUSY-breaking mass to yield the physical tree-level sfermion mass of zero once the cosmological constant is tuned to zero.  This interpretation is probably too pessimistic, though, since the tachyonic uplifting is an automatic consequence of adjusting the cosmological constant.  Concretely, this uplifted mass arises from the scalar auxiliary field (and the corresponding goldstino couplings arise from mixing with the gravitino), so once you have the sequestered form of $K$ and $W$, you necessarily obtain zero scalar masses but non-zero $a_{i \j}$ couplings. 

A second, more optimistic, interpretation is that \Eq{eq:irreduciblegoldstino} gives a concrete definition of sequestering.  While the extra-dimensional picture in \Fig{fig:sequestered} is a nice realization of sequestering, the sequestered limit can be achieved in more general theories.  In four-dimensional models with conformal sequestering \cite{Luty:2001jh,Luty:2001zv,Schmaltz:2006qs}, the visible and hidden sectors effectively decouple under RG flow to the infrared, assuming all composite vector multiplets in the hidden sector have mass dimension greater than 2.  As we explain in \App{sec:RGstability}, \Eq{eq:irreduciblegoldstino} is actually RG invariant, so one might conjecture that it corresponds to precisely the (attractive) IR fixed point needed to have a conformally sequestered theory.  More generally, one can identify when a theory is sequestered if \Eq{eq:irreduciblegoldstino} (and corresponding loop corrections, see \Sec{subsec:finalanswer}) is the only coupling between the visible and hidden sectors.\footnote{As shown in \Ref{D'Eramo:2012qd}, the sequestered limit implies that gaugino-gauge boson-goldstino couplings are zero.}

Regardless of how one interprets this result, the irreducible goldstino coupling is an unavoidable consequence of AdS SUSY lifted to flat space, since something needs to lift the tachyonic scalars to have a stable theory in flat space.  One might even hope to measure \Eq{eq:irreduciblegoldstino} experimentally as a way to gain access to the underlying AdS curvature.  

\subsection{Giudice-Masiero Terms}

The Giudice-Masiero mechanism \cite{Giudice:1988yz} is a way to generate a $\mu$ term and a $B_\mu$ term proportional to $m_{3/2}$ without (apparently) requiring couplings between the visible and hidden sectors.  Via a holomorphic piece in the K\"ahler potential (written using boldface to emphasize that these are superfields)
\be
\label{eq:GMkahler}
-3 e^{-3\bK} \supset \epsilon \bH_u \bH_d + \text{h.c.},
\ee
one generates the fermion and scalar mass terms
\be
\mathcal{L} \supset - \epsilon m_{3/2} \psi_u \psi_d - \epsilon m_{3/2}^2 h_u h_d + \text{h.c.}  \quad \Rightarrow \quad\frac{B_\mu}{\mu} = + m_{3/2}.
\ee
The sign of $B_\mu$ here is crucial, since if instead one had the superpotential
\be
\bW \supset \mu \bH_u \bH_d,
\ee
the fermion and scalar mass terms would be
\be
\mathcal{L} \supset - \mu \psi_u \psi_d + m_{3/2} \mu h_u h_d + \text{h.c.}  \quad \Rightarrow \quad \frac{B_\mu}{\mu} = - m_{3/2}.
\ee
From \Eq{eq:goldstinocouplings}, we see that the Giudice-Masiero mechanism actually does break SUSY (with $b_{ij} = 2 m_{3/2} \mu$), while generating $B_\mu$ from the superpotential does not break SUSY (i.e.~$b_{ij} = 0$).  Written in this language, it is confusing how a goldstino coupling could appear in the Giudice-Masiero mechanism since there is no goldstino present in \Eq{eq:GMkahler}.  

We can do a K\"ahler transformation to make the physics manifest.  To model SUSY breaking, we use a non-linear goldstino multiplet~\cite{Rocek:1978nb,Lindstrom:1979kq,Komargodski:2009rz,Cheung:2010mc,Cheung:2011jq}
\be
\bX_{\NL} = F_X \left( \theta+ \frac{1}{\sqrt{2} F_X}  \gold   \right)^2
\label{eq:XNL}
\ee
that satisfies $\bX_{\NL}^2 = 0$.  In a theory where the visible Higgs multiplets are sequestered from SUSY-breaking, the relevant pieces of the K\"ahler potential and superpotential are
\begin{align}
-3 e^{-\bK/3} & = -3 + \bX_{\NL}^\dagger \bX_{\NL} + \epsilon ( \bH_u \bH_d + \text{h.c.}) + \ldots, \label{eq:GMsequestered}\\
\bW & = m_{3/2} + f \bX_{\NL}  + \ldots,
\end{align}
where the equations of motion set $F_X^* = -f$ and fine-tuning the cosmological constant to zero requires $f = \sqrt{3} m_{3/2}$.  At tree-level, the physics is invariant to doing a K\"ahler transformation\footnote{At loop level, one must account for the K\"ahler anomaly \cite{Bagger:1999rd}.}
\be
\bK \to \bK + \bOmega + \bOmega^\dagger, \qquad \bW \to e^{-\bOmega} \bW,
\ee
so choosing $\bOmega = - \epsilon \bH_u \bH_d$, we have
\begin{align}
-3 e^{-\bK/3} & = -3 + \bX_{\NL}^\dagger \bX_{\NL} - \frac{\epsilon}{3} \bX_{\NL}^\dagger \bX_{\NL}(\bH_u \bH_d  + \text{h.c.})  + \ldots, \\
\bW & = m_{3/2} + f \bX_{\NL} + \epsilon m_{3/2}  \bH_u \bH_d + \epsilon f  \bX_{\NL}  \bH_u \bH_d + \ldots.
\end{align}
We see immediately that the Higgs multiplets have a SUSY-preserving $\mu = \epsilon m_{3/2}$, and a corresponding SUSY-preserving contribution to $B_\mu$ of $-\mu m_{3/2} = -\epsilon m_{3/2}^2$.  But there are also SUSY-breaking $B_\mu$ terms from direct couplings to $\bX_{\NL}$ in both the K\"ahler potential and superpotential.  This yields a contribution to $B_\mu$ of $(-\frac{1}{3} + 1) \epsilon |f|^2$, which equals $+2 \epsilon m_{3/2}^2$ after tuning the cosmological constant to zero.  Therefore, we have
\be
B_\mu = -\epsilon m_{3/2}^2 + 2 \epsilon m_{3/2}^2 = +\epsilon m_{3/2}^2, \qquad b_{ij} =  2 \epsilon m_{3/2}^2,
\ee
as required by \Eq{eq:goldstinocouplings}.  

Despite the fact that Giudice-Masiero can be written in a sequestered form in \Eq{eq:GMsequestered}, there is secretly a coupling between the visible sector Higgs multiplets and the hidden sector goldstino.\footnote{Of course, the physics is invariant to K\"ahler transformations at tree-level; all we have done here is choose a convenient K\"ahler basis to make the physics more clear.}  Thus, we conclude that the relation ${B_\mu}/{\mu} = + m_{3/2}$ is due to a partial cancellation between a SUSY-preserving and a SUSY-breaking effect, and corresponds to a tuning between (otherwise) independent parameters.  In the strict sequestered limit where only irreducible goldstino couplings are allowed, Giudice-Masiero terms must be absent.  

\subsection{Mass Splittings for Regulators}
\label{sec:regulatorsplittings}

In order to set the stage for talking about anomaly mediation at loop level in the next section, we want to discuss a bit about the physics that regulates logarithmic UV divergences in SUGRA.  There are various ways to introduce an effective cut-off scale $\Lambda_{\rm UV}$ into SUGRA, for example by introducing Pauli-Villars regulators \cite{Gaillard:1993es,Gaillard:1996ms} or higher-dimension operators that regulate the UV behavior \cite{Kaplunovsky:1994fg}.  However, already at tree-level, we can see the consequences of having a physical regulator in AdS SUSY.

Consider a Pauli-Villars chiral regulator field with a SUSY-preserving mass $\Lambda_{\rm UV}$.  If this regulator does not break AdS SUSY, then it must have an additional scalar negative mass-squared $-2m_{3/2}^2$ as well as a $B$-term of $-m_{3/2} \Lambda_{\rm UV}$, giving rise to SUSY-preserving mass splittings between the Pauli-Villars fermions and scalars:
\be
\label{eq:regulatormasses}
m^2_{ \text{PV-scalar}} = - 2 m_{3/2}^2 + \Lambda^2_{\rm UV} \pm m_{3/2} \Lambda_{\rm UV} , \qquad m_{\text{PV-fermion}} = \Lambda_{\rm UV}.
\ee
Any UV-divergent SUGRA calculation that properly includes the regulator modes will be affected by this mass splitting, and this fact is one way to understand the necessity of anomaly mediation.\footnote{In \Sec{sec:effectofregulators}, we will show how the regulators must be included to get super-Weyl-invariant gaugino masses.}  We often say that anomaly mediation is ``gauge mediation by the regulators'', in the sense that the (SUSY-preserving) mass splitting at the threshold $\Lambda_{\rm UV}$ acts analogously to the (SUSY-breaking) messenger mass threshold of gauge mediation.  Crucially, we will see that the mass splittings generated by anomaly mediation do not break AdS SUSY.

It is possible, however, to regulate SUGRA with a regulator multiplet whose scalar and fermionic components have a common mass $\Lambda_{\rm UV}$, for example by appropriately coupling the regulators to the SUSY-breaking $\bX_{\NL}$.  All this means is that the regulator multiplet must have corresponding goldstino couplings by conservation of the AdS supercurrent:  
\be
\label{eq:regulatorgoldstino}
a_{\text{PV}} = 2 m_{3/2}^2, \qquad b_{\text{PV}} =  m_{3/2} \Lambda_{\rm UV}.
\ee
Since there is no mass splitting among the regulators, no mass splittings are generated.  However, we would instead get goldstino couplings from the regulator fields!  One can of course consider an intermediate case with a combination of mass splittings and goldstino couplings.  In either event, one can show that modifying regulator couplings in this fashion is phenomenologically equivalent to changing $\vev{K_i F^i}$ for the purposes of loop-level calculations,\footnote{In the language of \Sec{sec:oneloop}, coupling regulators in such a fashion is largely equivalent to making the replacement $\bC \rightarrow \bC (1 + \bX_{\rm NL}/\Lambda)$, with $\bC$ the Weyl compensator.} so for simplicity we will assume regulators have no explicit coupling to SUSY breaking in the subsequent sections.\footnote{To avoid later confusion, we want to point out that there are two different types of ambiguities.  The ambiguity discussed here is whether the regulators do or do not experience SUSY breaking, which is a physical effect that can be measured using goldstino couplings.  There is a separate ambiguity in \Sec{subsec:superbox} having to do with how to write down a SUGRA-invariant 1PI effective action.  This is (partially) resolved using supertraces to define the soft mass spectrum, up to a puzzling ambiguity in how the $c_7$ term affects $\mathcal{T}$.}

\section{Anomaly Mediation and Super-Weyl Invariance}
\label{sec:oneloop}

In \Ref{D'Eramo:2012qd}, we described one-loop anomaly-mediated gaugino masses using the conformal compensator formalism of SUGRA \cite{Siegel:1978mj,Kugo:1982cu,Gates:1983nr}, which is a gauge fixing of super-conformal SUGRA.  Here, we will instead use the super-Weyl invariant formulation of SUGRA, which will allow us to connect directly to the claims of de Alwis in \Refs{deAlwis:2008aq,deAlwis:2012gr}.  Starting with a review of the super-Weyl formalism, we will follow the logic of de Alwis (which itself follows the logic of Kaplunovsky and Louis \cite{Kaplunovsky:1994fg}) to construct a Wilsonian effective action.  After demonstrating the existence of anomaly mediation in the Wilsonian picture, we derive the same effect using a super-Weyl invariant and SUSY-preserving 1PI effective action.  We will only consider gaugino masses in this section, leaving our main result on sfermion masses to \Sec{sec:twoloop}.

\subsection{Super-Weyl Formalism for SUGRA}
\label{eq:SWtreelevel}

The SUGRA lagrangian can be derived from a gauge fixing of super-Weyl-invariant SUGRA.  Super-Weyl transformations are the most general transformations that leave the torsion constraints of SUGRA unchanged, and they may be parameterized by a chiral superfield $\bSigma$ (and its conjugate anti-chiral superfield $\bSigma^\dagger$) \cite{Howe:1978km,wess1992supersymmetry}.  The components of the chiral superfield $\bSigma$ correspond to different types of transformations which may be familiar from the superconformal algebra: $\textrm{Re } \bSigma |$ corresponds to dilatations, $\textrm{Im } \bSigma|$ to chiral $U(1)_R$ rotations, and $\D_\alpha \bSigma |$ to conformal supersymmetry.  The $F_\Sigma$ component of $\bSigma$ corresponds to a new symmetry which will play a key role in understanding anomaly mediation.\footnote{Super-Weyl transformations do not include special conformal transformations, and superconformal transformations do not include the symmetry generated by $F_\Sigma$, so neither super-Weyl transformations nor superconformal transformations are a subset of the other.}    

The complete super-Weyl transformations are given in \App{app:superweyl}.  Crucially, the only field that transforms under $F_\Sigma$ is the scalar auxiliary field $M$ of supergravity \cite{Howe:1978km,wess1992supersymmetry,Kaplunovsky:1994fg}:
\begin{align}
\label{eq:Mtransform}
M^* & \rightarrow M^* - 6 F_\Sigma.
\end{align}
This auxiliary field appears in the determinant of the SUSY vielbein $\bE$,  the corresponding chiral density $2 \bEpsilon$, and chiral curvature superfield $\bR$:
\be
\bE \supset - \frac{1}{3} M^* \Theta^2 + \textrm{h.c.} + \frac{1}{9} |M|^2 \Theta^4 , \qquad 2 \bEpsilon \supset - e M^* \Theta^2, \qquad \bR \supset - \frac{1}{6} M  - \frac{1}{9} |M|^2 \Theta^2 + \ldots
\ee
We will often talk about the Weyl weights $w$ of chiral superfields $\bQ_w$ and vector superfields $\bV_w$ which transform as \cite{wess1992supersymmetry} 
\begin{align}
\bQ_w \to \bQ_w e^{w \bSigma}, \qquad \bV_{\! w} \to \bV_{\! w} \, e^{w (\bSigma + \bSigma^\dagger)}.
\end{align} 
Ordinary matter fields have Weyl weight 0, so the K\"ahler potential $\bK$ and superpotential $\bW$ also have Weyl weight 0.  For a vector superfield of weight 0, the gauge-covariant superfield $\bW_{\! \alpha}$ has Weyl weight $-3$.  In the gravity multplet, $\bE$ has Weyl weight 4 and $2 \bEpsilon$ has Weyl weight 6.

The usual SUGRA action (e.g.~in \Ref{wess1992supersymmetry}) is not invariant under super-Weyl transformations, so one needs to introduce a super-Weyl compensator $\bC$ with Weyl weight $-2$ (i.e.~$\bC \to e^{-2 \bSigma} \bC$).  In that case, the tree-level lagrangian
\begin{align}
\mathcal{L} & = \int d^4 \Theta \, \bE \, \bC^\dagger \bC \, (-3 e^{-\bK/3} ) + \int d^2 \Theta \, 2 \bEpsilon \, \bC^3 \bW + \frac{1}{4} \int d^2 \Theta \, 2 \bEpsilon \, \bW^\alpha \bW_{\! \alpha} + {\rm h.c.}  \label{eq:superweyllagrangian}
\end{align}
has Weyl weight 0 as desired.  The components of the super-Weyl compensator are
\be
\bC =  C \{ 1, \chi_C , F_C \},
\ee  
and due to the non-vanishing Weyl weight of $\bC$, $F_C$ transforms under $F_\Sigma$ as
\begin{align}
\label{eq:FCtransform}
F_C & \rightarrow F_C - 2 F_\Sigma.
\end{align} 

It should be stressed that this super-Weyl invariance (and the corresponding super-Weyl compensator) were introduced into \Eq{eq:superweyllagrangian} simply for calculational convenience, and physical results will not actually exhibit super-Weyl symmetry.  After all, one can use the super-Weyl transformations to gauge-fix $\bC$ in some convenient fashion, leaving a theory without spurious symmetries or degrees of freedom.  Because $F_\Sigma$ transformations are a gauge redundancy of the theory, though, physical observables will only depend on the combination\footnote{The superconformal formalism does not contain $M^*$, since that degree of freedom is contained in the $F_\Phi$ component of the conformal compensator (see \App{app:compensator}).  In the super-Weyl case, the $F_C$ component is a pure gauge degree of freedom.}
\be
\label{eq:FSW}
\FSW \equiv F_C - \frac{1}{3} M^*,
\ee
regardless of what gauge choice is ultimately made.  As we will argue, this $F_\Sigma$-invariance is the key point missed in \Refs{deAlwis:2008aq,deAlwis:2012gr} (and implicitly missed in \Ref{Kaplunovsky:1994fg}).

\subsection{Choice of Gauge Fixing}

To recover the familiar SUGRA lagrangian from \Eq{eq:superweyllagrangian}, one must gauge fix $\bC$.  The choice $\bC = 1$ yields the lagrangian in ``SUGRA frame'' (i.e.~without performing any super-Weyl transformations).   A more convenient choice is \cite{Kaplunovsky:1994fg}
\begin{align}
\log \bC + \log \bC^\dagger & = \frac{1}{3} \bK|_H, \label{eq:superweylgaugefixing}
\end{align}
with $\bK|_H$ being the harmonic (i.e.~chiral plus anti-chiral) part of the K\"ahler potential. This yields the lagrangian in ``Einstein frame'' (i.e.~after having performed appropriate super-Weyl transformations).   Effectively, this gauge choice is the equivalent of going to Wess-Zumino gauge for the real superfield $\bK$.\footnote{This gauge choice leaves still leaves $\arg C$ undetermined, though one can fix $\arg C$ by imposing  that the gravitino mass parameter has no phase.}  It must be stressed that \Eq{eq:superweylgaugefixing} is not a supersymmetric relation amongst superfields, since $\bK|_H$ is not a superfield itself.  Instead, \Eq{eq:superweylgaugefixing} should be thought of merely as a prescription for setting each component of $\bC$ and $\bC^\dagger$.   Of course, other gauge-fixing prescriptions will give physically equivalent results, but \Eq{eq:superweylgaugefixing} is particularly convenient since this choice for $\Re C$ yields canonically-normalized Einstein-Hilbert and Rarita-Schwinger terms and this choice for $\chi_C$ eliminates troublesome matter-gravitino mixings.

However, it is not so clear what is accomplished by gauge-fixing $F_C$.  We can investigate this by examining the portion of \Eq{eq:superweyllagrangian} that depends on $F_C$ and $M^*$, since these are the only two fields that are not inert under $F_\Sigma$ transformations.  
\begin{align}
e^{-1} \mathcal{L} & =C^* C \left(e^{-K/3} \right) \left( - 3 \left(F_C^* - \frac{1}{3} M \right) \left( F_C - \frac{1}{3} M^* \right) + K_i F^i \left(F_C^* - \frac{1}{3} M \right)  + \textrm{h.c.}  \right)\nn \\
& \quad \, + 3 C^3 \left(F_C - \frac{1}{3} M^* \right) W + \textrm{h.c.} + \ldots %
\label{eq:FCandM}
\end{align}
As expected from \Eq{eq:FSW}, $F_C$ and $M^*$ only appear in the $F_\Sigma$-invariant combination $\FSW \equiv F_C - \frac{1}{3} M^*$ which has the vacuum expectation value
\begin{align}
\vev{\FSW} = m_{3/2} + \frac{1}{3} \vev{K_i F^i}.
\label{eq:FSWvev}
\end{align}
Thus, different gauge-fixings for $F_C$ only serve to shift the vev of $M^*$.  After one solves the $M^*$ equation of motion, physical observables do not (and cannot) depend on the gauge fixing of $F_C$.

\subsection{Counterterms in the Wilsonian Effective Action}

As emphasized in \Ref{deAlwis:2008aq,deAlwis:2012gr}, it is possible to regulate all UV-divergences in SUGRA in a way that preserves SUSY and super-Weyl invariance.   This was shown in \Ref{Kaplunovsky:1994fg} using higher-derivative regulators in a version of Warr's regularization scheme \cite{Warr:1986we,Warr:1986ux}.  This implies that the super-Weyl symmetry discussed above is not anomalous, and consequently, any physical results we derive must be completely super-Weyl invariant.  Indeed, we will see that anomaly mediation (despite its name) is necessary to preserve both SUSY and super-Weyl invariance.

The key observation of \Ref{Kaplunovsky:1994fg} is that to preserve super-Weyl invariance in a UV-regulated theory, the Wilsonian effective action must consist of \Eq{eq:superweyllagrangian} augmented with the counterterm 
\be
\Delta \mathcal{L}  = \frac{3}{16 \pi^2} (T_G - T_R) \int d^2 \Theta \, 2 \bEpsilon \, \log \bC \, \bW^\alpha_{\! a} \bW^a_{\! \alpha}. \label{eq:fixinganomaly}
\ee
This term can be deduced from the requirement that the $U(1)_R$ part of the super-Weyl transformations remains non-anomalous.  It is convenient to canonically normalize the matter fields $\bQ^i$ by performing the (anomalous) rescaling $\bQ^i \rightarrow  \bQ^i / \bC$ such that the rescaled matter field have Weyl weight $-2$.  Due to the Konishi anomaly \cite{Konishi:1983hf,Clark:1979te}, this rescaling modifies \Eq{eq:fixinganomaly} to become
\be
\label{eq:finalfixinganomaly}
\Delta \mathcal{L}  = \frac{1}{16 \pi^2} (3T_G - T_R) \int d^2 \Theta \, 2 \bEpsilon \, \log \bC \, \bW^\alpha_{\! a} \bW^a_{\! \alpha}.
\ee
Immediately this presents a conundrum, since \Eq{eq:finalfixinganomaly} contains a gaugino mass that depends only on $F_C$:
\be
m^{\rm ambiguous}_\lambda = - \frac{g^2}{16 \pi^2} \left(3 T_G - T_R \right) F_C .
\label{eq:gauginomassFC}
\ee
Following the analysis of \Ref{Kaplunovsky:1994fg}, \Refs{deAlwis:2008aq,deAlwis:2012gr} claimed this was the complete formula for the gaugino mass, and by gauge-fixing $F_C = \frac{1}{3}K_i F^i$ as in \Eq{eq:superweylgaugefixing}, de Alwis found no contribution to $m_\lambda$ proportional to the gravitino mass $m_{3/2}$, and hence no anomaly mediation.\footnote{In the language of \Ref{D'Eramo:2012qd}, de Alwis was only claiming the absence of gravitino mediation.  The K\"ahler-mediated terms proportional to $K_i F^i$ are not in dispute.} 

However, we see immediately that \Eq{eq:gauginomassFC} cannot be the complete answer, since it is not invariant under $F_\Sigma$ transformations.  This is incompatible with the assertion that the physical predictions of this theory should be invariant under such super-Weyl transformations.  By \Eq{eq:FSW}, the physics should depend on the combination $\FSW \equiv F_C - \frac{1}{3} M^*$ (which does contain $m_{3/2}$).  One could try to make the replacement 
\be
\log \bC \to \log \bC + \frac{1}{3} \log 2 \bEpsilon
\ee
to make the dependence on $\FSW$ manifest, but as emphasized emphatically (and correctly) in \Refs{deAlwis:2008aq,deAlwis:2012gr}, $2 \bEpsilon$ is a chiral density and not a chiral superfield, and one cannot include arbitrary extra factors of a chiral density in a SUGRA-invariant action, just as one cannot include arbitrary extra factors of $\det e$ in a diffeomorphically-invariant action.  Indeed, there is no local term that one can add to the Wilsonian action to make \Eq{eq:finalfixinganomaly} manifestly super-Weyl invariant.\footnote{We will see in \Sec{subsec:1PI} that one can write down a non-local 1PI effective action that depends only on $\FSW$.}

\subsection{Effect of the Regulators}
\label{sec:effectofregulators}

The resolution to the above puzzle is that the Wilsonian effective action (as defined in \Ref{Kaplunovsky:1994fg}) \emph{needs} to violate super-Weyl invariance in order for physical results to be super-Weyl invariant.  This is familiar from Yang-Mills gauge theories with a hard Wilsonian cutoff, where the Wilsonian action must be non-gauge invariant in order compensate for the non-gauge invariance of the cutoff (see also \Ref{Dine:2007me}).  In this case, the tree-level expression in \Eq{eq:gauginomassFC} will combine with loops of the regulators to yield a super-Weyl invariant result.

To understand how this effect arises, consider a Pauli-Villars regulator, as anticipated in \Sec{sec:regulatorsplittings}.  Given a chiral superfield $\bQ$ in some representation of a gauge group, one can regulate its contributions to loop diagrams by introducing two superfields, $\bL$ and $\bS$, with $\bL$ in the same representation of the gauge group and $\bS$ in the conjugate representation:  
\be
\mathcal{L}_{\rm PV} =  \int d^4 \Theta \, \bE \, \left[- {\bL}^\dagger e^{\bV} {\bL} -  {\bS}^\dagger e^{\bV} {\bS} \right]  + 
\int d^2 \Theta \, 2 \bEpsilon \, \Lambda_{\rm PV} \, {\bf L} \, {\bf S} + {\rm h.c.} 
\label{eq:PV}
\ee
%Here, we take the regulator have already performed the rescaling such that all matter and regulator fields have Weyl weight $-2$. \fde%{I am confused about the rescaling of the regulator fields (for matter fields of course is the usual story). What do we mean when we say that we did the rescaling? If we did something analogous to matter fields, my understanding is that the superpotential term before the rescaling was $C^2 W$, such that the $C$ disappears. I guess the question is: if there is a rescaling, what was the Lagrangian before the rescaling? I have no doubt Eq.(3.15) is correct, but I think we should explain better how we get it.} 
Gauge fields can be similarly regulated by introducing chiral superfield regulators in the adjoint representation.  By using many such regulators and including appropriate couplings, all divergences of SUGRA can be removed \cite{Gaillard:1994mn,Gaillard:1993es,Gaillard:1996ms}.  The kinetic terms suggest that the regulator fields have Weyl weight $-2$, but since the Pauli-Villars mass term is $\Lambda_{\rm PV}$ instead of $\bC \Lambda_{\rm PV}$, the Pauli-Villars fields break super-Weyl invariance.   However, \Ref{Kaplunovsky:1994fg} showed that \Eq{eq:finalfixinganomaly} is precisely the term needed to restore super-Weyl invariance of the action.

Now, because the Pauli-Villars regulators have a SUSY-preserving mass $\Lambda_{\rm PV}$, they exhibit boson/fermion mass splitting due to the $\Theta^2$ component of $2\bEpsilon$.  Expanding \Eq{eq:PV}, we find
\be
\mathcal{L}_{\rm PV}  \supset - \frac{1}{3} \Lambda_{\rm UV}M^*  \, L \, S \, ,
\ee
which is a $B$-term that is not super-Weyl invariant!  Doing calculations with these regulators will yield an $M^*$-dependent gaugino mass at one loop.   Adding this loop-level contribution to the tree-level contribution from \Eq{eq:gauginomassFC}, we have the super-Weyl invariant gaugino mass
\be
\label{eq:correctgaugino}
m^{\rm physical}_\lambda = - \frac{g^2}{16 \pi^2} \left(3 T_G - T_R \right) \FSW  = - \frac{g^2}{16 \pi^2} \left(3 T_G - T_R \right) \left(m_{3/2} + \frac{1}{3} K_i F^i   \right).
\ee
This expression is manifestly super-Weyl invariant, and reproduces the familiar anomaly-mediated result.  As discussed in \Sec{sec:regulatorsplittings}, if the regulators couple to SUSY breaking in such a way to remove the  $m_{3/2}$ dependence in the gaugino mass, this effect would show up as an $m_{3/2}$ dependence in the associated goldstino couplings.

One can avoid this subtlety of regulator contributions by making a gauge choice such that the vev $\vev{M^*} = 0$.  In that gauge (and only for that gauge), there are no regulator $B$-terms, so \Eq{eq:gauginomassFC} then yields the correct gaugino mass with $F_C = m_{3/2} + \frac{1}{3} K_i F^i$.\footnote{It is worth noting here that $m_{3/2}$ here is really the vev of the superpotential $W$, which is allowed to appear in the gauge fixing of $F_C$.}  This is essentially the strategy used in \Ref{D'Eramo:2012qd} (since the superconformal framework automatically sets $M^* = 0$), and is effectively what was done in the original anomaly-mediated literature \cite{Randall:1998uk,Giudice:1998xp} (though not in this language).  For any other gauge---including the choice of \Eq{eq:superweylgaugefixing} used by \Refs{Kaplunovsky:1994fg,deAlwis:2008aq,deAlwis:2012gr}---one cannot neglect contributions to the gaugino mass due to the UV regulators.  Alternatively, one can regulate the theory with super-Weyl-invariant Pauli-Villars fields, in which case \Eq{eq:finalfixinganomaly} is absent but the regulators have $B$-terms proportional to $\FSW$, again reproducing \Eq{eq:correctgaugino}.

 \subsection{1PI Effective Action and Goldstino Couplings}
\label{subsec:1PI}

We argued above that there is no way to make super-Weyl invariance manifest in a Wilsonian effective action.  However, the super-Weyl formalism is entirely valid at the quantum level, since there exists a variety of regularization schemes that preserve the super-Weyl symmetry (i.e.~it is not anomalous).  Therefore, we should be able to write down a 1PI effective action that exhibits all of the relevant symmetries of the theory (including super-Weyl invariance).  Here, we will write down the relevant 1PI action to describe gauginos at one loop, and extend the logic to sfermions at two loops in \Sec{sec:twoloop}.  

One disadvantage of the 1PI action is that it will inevitably be non-local, since it involves integrating out light degrees of freedom.  On the other hand, the 1PI action allows us to extract all anomaly-mediated effects from the action directly, without having to worry about the contributions of regulators explicitly as we did in \Sec{sec:effectofregulators}.  To avoid SUSY-breaking terms in the regulators as discussed in \Sec{sec:regulatorsplittings}, we can study a 1PI effective action that does not have explicit dependence on $\bX_{\rm NL}$.  In general, the 1PI effective action will depend on $\bX_{\rm NL}$, but this will just give extra soft masses and goldstino couplings in agreement with flat space intuition, whereas we are interested in isolating the anomaly-mediated effects.

At one-loop, the 1PI effective action for the gauge multiplet is
\begin{align}
\mathcal{L} & \supset \frac{1}{4} \int d^2 \Theta \, 2 \bEpsilon \bW^\alpha \bS (\superbox) \bW_{\! \alpha}, \label{eq:chiral1PI}
\end{align}
The superfield $\bS$ is a chiral superfield with the gauge coupling as its lowest component (see \Ref{ArkaniHamed:1998kj}).  The running of the coupling with the momentum scale is encapsulated by the dependence of $\bS$ on $\superbox$, an appropriately SUGRA-covariant, super-Weyl-covariant, and chiral version of the d'Alembertian.  This 1PI action depends on the holomorphic gauge coupling, which is sufficient if we are only interested in one-loop expressions.  To describe the canonical gauge coupling (including two-loop effects), one needs an alternative action described in \App{sec:1PIgaugino}.

As we will discuss further in \Sec{sec:twoloop}, the choice of $\superbox$ is in fact ambiguous.  All choices are equivalent at $\mathcal{O}(m_{3/2})$, though, and we will choose to work with\footnote{\Ref{Kaplunovsky:1994fg} never explicitly wrote down the form for $\superbox$ acting on $\bW_{\! \alpha}$.  This slightly complicated form is needed because $\bW_{\! \alpha}$ has a spinor index.}
\be
\label{eq:superboxfirstdef}
\superbox \bW_{\! \alpha} \equiv -\frac{1}{8} (\D^{\dagger 2} - 8 \bR)  \D_\alpha \left[ \frac{\D^\beta \bW_{\! \beta}}{\bC^\dagger \bC} \right].
\ee
It is then possible to expand out \Eq{eq:chiral1PI} and derive super-Weyl-invariant gaugino masses and goldstino couplings.\footnote{As written, this form of $\superbox$ is only gauge-invariant for an abelian gauge symmetry.  It can be easily modified for non-abelian gauge symmetries by appropriate insertions of $e^{\pm \bV}$.} Note that $\superbox \bW_{\! \alpha}$, like $\bW_{\! \alpha}$, is chiral and has Weyl weight $-3$.  

In practice, though, it is much more convenient to use the $F_\Sigma$ gauge freedom to set $M^* = 0$.  The remaining components of $\bC$ can be fixed using the gauge choice in \Eq{eq:superweylgaugefixing} such that (to linear order in fields)
\be
\bC = \left\{1, \frac{1}{3} K_i  \chi^i , m_{3/2} + \frac{1}{3} K_i  F^i \right\}.  \label{eq:m32gaugechoice}
\ee
Note that the fermionic component of $\bC$ contains a goldstino if $K_i$ attains a vev: 
\be
\chi_C  = \frac{1}{3} \vev{K_i F^i} \frac{\gold}{F_{\rm eff}}.
\ee
In this gauge, the graviton and gravitino are canonically normalized and there are no gravitino-goldstino kinetic mixing terms to worry about.  We can also drop the chiral curvature superfield $\bR$ in \Eq{eq:superboxfirstdef} because it only contributes at $\mathcal{O}(m_{3/2}^2)$ in $M^* = 0$ gauge (and in fact gives no contribution in this gauge if the cosmological constant has been tuned to zero).  Similarly, $-\frac{1}{8} \D^{\dagger 2} \D_\alpha \D^\beta \bW_{\! \beta}$  equals the ordinary flat space d'Alembertian $\square$ acting on $\bW_{\! \alpha}$ at this order.  So for the purposes of getting the $\mathcal{O}(m_{3/2})$ gaugino mass and goldstino couplings, we can simply make the replacement
\be
\label{eq:easychange}
\superbox \to \frac{1}{\bC^\dagger \bC} \square + \frac{1}{2} i (\D^\dagger_{\alphadot} \bC^\dagger) \sigmabar^{\mu \alphadot \beta} \partial_\mu \D_\beta - \frac{1}{16} \left(\D^{\dagger 2} \bC^\dagger \right) \D^2,
\ee
where we have dropped terms with superspace derivatives on multiple copies of $\bC$ (they never contribute at $\mathcal{O}(m_{3/2})$) and terms with spacetime derivatives on $\bC$ (they would only yield terms with derivatives on goldstinos, which can be ignored at this order in $m_{3/2}$ in the goldstino equivalence limit).  The form of $\superbox$ in \Eq{eq:easychange} is not as manifestly chiral as in \Eq{eq:superboxfirstdef}, but it can be verified to be chiral (up to terms that we have dropped at this order).

This gauge choice for $\bC$ is equal to the gauge choice for the conformal compensator $\bPhi$ used in \Ref{D'Eramo:2012qd}, and yields identical results.  Plugging \Eq{eq:easychange} into \Eq{eq:chiral1PI} yields the expected soft masses and goldstino couplings from traditional anomaly mediation:\footnote{Strictly speaking, this is only the piece of anomaly mediation related to the super-Weyl anomaly.  See \Refs{Bagger:1999rd,D'Eramo:2012qd} for how the K\"ahler and Sigma-Model anomalies contribute to the 1PI effective action.} 
\be
\label{eq:genericGoldstinoCouplings}
\mathcal{L} \supset - \frac{1}{2}m_{\lambda} \lambda_a \lambda^a + \frac{c_\lambda}{\sqrt{2} F_{\rm eff}} \lambda_a \sigma^{\mu \nu} \gold F_{\mu \nu}^a,
\ee
where
\be
m_\lambda = - \frac{\beta_g}{g} \left(m_{3/2} +\frac{1}{3} K_i F^i \right), \qquad c_\lambda = -\frac{\beta_g}{g} \frac{1}{3} K_i F^i,
\ee
and $\beta_g$ is the beta function for the relevant gauge group.  Note that the piece of $m_\lambda$ proportional to $m_{3/2}$ does not come with a goldstino coupling, which tells us that it is not an (AdS) SUSY breaking effect.  Had we instead worked in a gauge where $M^* = -3m_{3/2}$ (as was the case in \Refs{deAlwis:2008aq,deAlwis:2012gr}), then the gaugino mass proportional to $m_{3/2}$ would arise from the parts of $\superbox$ that depend on the lowest component of the chiral curvature superfield $\bR$.

Thus, we have seen how anomaly mediation is a necessary consequence of SUSY invariance and super-Weyl invariance.  Because of the underlying AdS SUSY algebra, terms proportional to $m_{3/2}$ necessarily appear in the regulated SUGRA action.  Crucially, $m_{3/2}$ is not an order parameter for (AdS) SUSY breaking, so anomaly-mediated soft masses proportional to $m_{3/2}$ do not have associated goldstino couplings.

\section{All-Orders Sfermion Spectrum from Anomaly Mediation}
\label{sec:twoloop}

It is well-known that anomaly mediation yields sfermion soft mass-squareds at two loops proportional to $m_{3/2}^2$~\cite{Randall:1998uk}.  In this section, we want to show that this effect can be understood as being a consequence of AdS SUSY.  To do so, we will follow the logic of \Sec{subsec:1PI} and derive the sfermion spectrum by constructing a super-Weyl-invariant and SUSY-preserving 1PI effective action for chiral multiplets.  

The obvious choice for the 1PI effective action is 
\begin{align}
\mathcal{L} & = \int d^4 \Theta \, \bE \, \bC^\dagger \bC \, \bQ^\dagger \bZ(\superbox) \bQ . 
\label{eq:1PIscalaraction}
\end{align}
Here, $\bQ$ is a chiral matter multiplet, $\bZ$ is the superfield associated with wave function renormalization, and $\superbox$ is a super-Weyl invariant version of the d'Alembertian acting on chiral superfields.  Our key task in this section is to figure out which pieces of \Eq{eq:1PIscalaraction} preserve SUSY and which pieces break SUSY.  To do this, we first identify the order parameter $F_R$ for SUSY breaking in the SUGRA multiplet, which is valid at order $\mathcal{O}(m_{3/2}^2)$.  We then use $F_R$ to help identify all places where the goldstino field can appear.  Because $\superbox$ is in fact ambiguous at $\mathcal{O}(m_{3/2}^2)$, we will need to construct appropriate supertraces to extract unambiguous ``soft mass-squareds'' and ``goldstino couplings''.  With these tools in hand, we can then use the 1PI effective action to derive the familiar two-loop scalar soft mass-squareds, as well as unfamiliar one-loop goldstino couplings.

\subsection{The Order Parameter for SUSY Breaking}

As already emphasized a number of times,  the gravitino mass  $m_{3/2}$ is not an order parameter for SUSY breaking but is simply a measure of the curvature of unbroken AdS space.  With an appropriate gauge choice (see \Eq{eq:linearCgauge} below), we can extract $m_{3/2}$ from the lowest component of the chiral curvature superfield $\bR$,
\be
\label{eq:lowestR}
\bR| = - \frac{1}{6} M^*= \frac{1}{2} m_{3/2} ,
\ee
and effects proportional to $\bR|$ will preserve (AdS) SUSY.

The SUGRA multiplet does contain a SUSY-breaking order parameter at order $\mathcal{O}(m_{3/2}^2)$, namely the highest component of $\bR$:
\be
- \frac{1}{4} \D^2 \bR | = \frac{1}{12} \mathcal{R} - \frac{1}{9} M^* M  + \ldots,
\ee
where $\mathcal{R}$ is the Ricci scalar.  Upon using the Einstein equation, this takes on the value
\begin{align}
F_R \equiv  \frac{1}{12} \mathcal{R}  - m_{3/2}^2 = - \frac{F_{\rm eff}^2}{3 \MPl^2} \ , \label{eq:FRvalue}
\end{align}
regardless of whether $F_{\rm eff}$ is tuned to yield flat space or not.  Since $F_{\rm eff}$ is an order parameter for SUSY breaking, so is $F_R$ for finite $\MPl$.  In an arbitrary gauge, we will define $F_R$ in terms of \Eq{eq:FRvalue} (instead of $- \frac{1}{4} \D^2 \bR |$).

As expected, $F_R = 0$ for unbroken AdS SUSY (i.e.~$\frac{1}{12}\mathcal{R} = m_{3/2}^2$).  When SUSY is broken and the cosmological constant is tuned to zero, then $F_R = - m_{3/2}^2$ (i.e.~$\mathcal{R} = 0$).  So while $m_{3/2}$ itself does not break SUSY, $F_R$ can yield effects proportional to $m_{3/2}^2$ that do break SUSY.  This distinction lies at the heart of the confusion surrounding anomaly mediation.

To better understand why $F_R$ is an order parameter for SUSY-breaking, it is helpful to note that $F_R$ controls the amount of gravitino-goldstino mixing in the super-Higgs mechanism.  This can be seen  by examining the various forms of the gravitino equation of motion one can obtain by plugging \Eq{eq:supercurrentWithGold} into \Eq{eq:gravitinoLag}: 
\begin{align}
\frac{1}{\MPl} \epsilon^{\mu \nu \rho \tau} \sigmabar_\nu \D_\rho \psi_\tau & = - \frac{3 i}{\sqrt{2}} \frac{F_R}{F_{\rm eff}} \sigmabar^\mu \gold + \ldots,  \nonumber \\ 
\frac{1}{\MPl} \sigma^{\mu \nu} \D_\mu \psi_\nu & = - \frac{3}{\sqrt{2}} \frac{F_R}{F_{\rm eff}} \gold + \ldots, \label{eq:gravitinoeoms} \\ 
- \frac{1}{\MPl} i \sigmabar^\mu \D_\mu \psi_\lambda & = \frac{3 i}{\sqrt{2}} \frac{F_R}{F_{\rm eff}} \sigmabar_\lambda \gold + \ldots, \nonumber
\end{align}
where we have also used the Einstein equation from \Eq{eq:FRvalue}. Thus, gravitino couplings which look innocuous can secretly contain (SUSY-breaking) goldstino couplings when $F_R$ is non-zero.  This will be of great importance when we track goldstino couplings in the next subsection. The ellipses of \Eq{eq:gravitinoeoms} contain terms not relevant to our discussion. In particular, we can ignore any $m_{3/2} \psi_\mu$ terms since we only care about effects up to $\mathcal{O}(m_{3/2}^2)$.  We can also ignore terms proportional to $\sigmabar^\mu \psi_\mu$, since applying its equation of motion would only serve to reintroduce derivatives acting either on gravitinos or goldstinos.
\subsection{Goldstinos in the SUGRA Multiplet}
\label{subsec:SUGRAmultiplet}

Since our ultimate goal is to compute the sfermion soft masses and goldstino couplings advertised in \Tab{tab:summary}, it is crucial to identify all places where the goldstino field can appear.

The most straightforward case is when there are direct couplings between the visible sector fields and the SUSY-breaking superfield $\bX_{\rm NL}$ from \Eq{eq:XNL}, which has the goldstino as its fermionic component.  This case is not interesting for our purposes since it generates soft masses and goldstino couplings in agreement with flat space intuition.  We therefore take the wavefunction superfield $\bZ$ to be independent of $\bX_{\rm NL}$ for simplicity.

Somewhat less obviously, the Weyl compensator $\bC$ itself can also contain a goldstino, and different (super-Weyl) gauge fixings give different goldstino dependence in $\bC$. We find it convenient to work in the gauge where 
\begin{align}
\bC & = \left\{1, \frac{1}{3} \vev{K_i}  \chi^i , \frac{1}{3} \vev{K_i}  F^i \right\}. \label{eq:linearCgauge}
\end{align}
This is effectively the gauge choice of \Eq{eq:superweylgaugefixing} carried out to linear order in fields, which is the minimum necessary to have canonically-normalized Einstein-Hilbert and Rarita-Schwinger terms \cite{Cheung:2011jp}. In this gauge $-\frac{1}{3} M^* = m_{3/2}$ (see \Eq{eq:lowestR}). Upon picking out the goldstino direction, neglecting other fermions, and dropping terms with multiple goldstinos,
\begin{align}
\bC & = 1 + \frac{1}{3} \vev{K_i F^i} \left( \Theta + \frac{\gold}{\sqrt{2} F_{\rm eff}} \right)^2 \ .
\end{align}
This gauge choice clearly shows that wherever $\vev{F_C} = \frac{1}{3} \vev{K_i  F^i}$ appears in a soft SUSY-breaking term, it will have an associated goldstino coupling. Of course, $F_C$ is always accompanied by $-\frac{1}{3} M^* = m_{3/2}$ by super-Weyl invariance, but effects proportional to $M^*$ do not have associated goldstino couplings.  After all, $\vev{M^*} \neq 0$ does not break AdS SUSY, whereas $\vev{K_i F^i} \neq 0$ does.
  
The most subtle case is to identify goldstino fields hiding in the SUGRA multiplet.  These arise through the gravitino equations of motion shown in \Eq{eq:gravitinoeoms}, which are necessarily SUSY invariant.  The SUSY transformation of \Eq{eq:gravitinoeoms} then tells us any goldstino arising in such a fashion must be accompanied by an $F_R$, thus giving us an easy way to track such goldstinos.  $F_R$ only occurs (without derivatives acting on it) within the SUGRA superfields $\bR$ and $\bG_\mu$, and the components of these superfields can be extracted by the methods of \Refs{wess1992supersymmetry,Baumann:2011nm}.\footnote{There are also goldstinos lurking in $\bE$, but these are most easily tracked by making the replacement
$$
\int d^4 \Theta \, \bE \, \bOmega = \frac{1}{2} \int d^2 \Theta \, 2 \bEpsilon \left[ - \frac{1}{4} (\D^{\dagger 2} - 8 \bR) \bOmega \right] + \rm h.c.,
$$
since $2 \bEpsilon$ does not have hidden goldstinos.}

Extensively using the gravitino equations of motion of \Eq{eq:gravitinoeoms}, we find that $\bR$ and $\bG_\mu$  can be written as:
\begin{align}
\bR & = - \frac{1}{6} M + F_R \left(\Theta + \frac{\gold}{\sqrt{2} F_{\rm eff}} \right)^2 + \ldots, \\
\bG_\mu & = \frac{1}{2} F_R \left(\Theta + \frac{\gold}{\sqrt{2} F_{\rm eff}} \right) \sigma_\mu \left(\Theta^\dagger + \frac{\gold^\dagger}{\sqrt{2} F_{\rm eff}} \right) + \ldots,
\end{align}
where the ellipses include terms containing $m_{3/2} \psi_\mu$, $\sigmabar^\mu \psi_\mu$, $b_\mu$, $\partial_\mu M$, $\partial_\mu F_R$,\footnote{Terms containing $\partial_\mu F_R$ (which has vanishing vev) may have associated goldstino couplings, but they will always feature a derivative acting on the goldstino.  Such terms will always be of $\mathcal{O}(m_{3/2}^3)$ in the goldstino equivalence regime, and can be ignored here.} or multiple gravitinos or goldstinos.  For simplicity, we have assumed that the Ricci tensor is proportional to the metric, as it is in any homogeneous space.

Note that with this particularly convenient gauge choice, we can identify all of the goldstino couplings in $\bX_{\rm NL}$, $\bC$, $\bR$, and $\bG_\mu$ by first finding the vevs of these fields, and then making the replacement
\be
\Theta \to \Theta + \frac{\gold}{\sqrt{2} F_{\rm eff}}. \label{eq:Thetareplacement}
\ee
At the component level, this implies that any terms in the lagrangian with coefficient $F_X$, $K_i F^i$, or $F_R$ (but crucially \emph{not} $m_{3/2}$) will have associated goldstino couplings.  These can be found by making a global SUSY transformation of those terms\footnote{The situation is more subtle for terms with coefficients like $m_{3/2} K_i F^i$, a product of SUSY-breaking and SUSY-preserving effects.  In such cases, one only makes half of the transformation of \Eq{eq:globalSUSYtransformation}.  This arises since for $K_i F^i$ ($K_\i F^{* \i}$), one is really only making the replacement of \Eq{eq:Thetareplacement} for $\Theta$ ($\Theta^\dagger$), not $\Theta^\dagger$ ($\Theta$), recalling that we have a hermitian action.} with infinitesimal SUSY parameter 
\begin{align}
\epsilon = - \frac{\gold}{\sqrt{2} F_{\rm eff} } .  \label{eq:globalSUSYtransformation}
\end{align}
This will allow us to identify goldstino couplings directly from the sfermion spectrum, without having to wrestle with complicated component manipulations.

The simplest application of this method for finding goldstino couplings is the tree-level analysis of \Sec{sec:treelevel}.  The tachyonic scalar masses are removed by a SUSY-breaking coupling $2 F_R \phi^* \phi$ when uplifting from AdS to flat space.  This indeed has a corresponding goldstino coupling in flat space proportional to $- 2 F_R / F_{\rm eff} = 2 m_{3/2}^2/ F_{\rm eff}$ (see \Eq{eq:irreduciblegoldstino}).\footnote{In practice, the use of gravitino equations of motion is less than transparent, which is the reason why we relied on the Einstein frame lagrangian in \Sec{subsec:SUGRAlagrangian}.  Finding the Einstein frame is more difficult beyond tree-level, however, which is why we choose to work in SUGRA frame in this section and exploit gravitino equations of motion.}

\subsection{Supertraces and the 1PI Effective Action}
\label{subsec:supertrace}

Now that we have identified our SUSY-breaking order parameters and how they are associated with goldstino couplings, we now need to consider what possible SUSY-breaking terms can arise from the 1PI effective action in \Eq{eq:1PIscalaraction}.  This action accounts for the quantum corrections coming from loop diagrams of massless particles. For this reason, one must be careful to include both local and non-local terms  when considering SUSY-breaking in a 1PI effective action.  For a chiral multiplet at quadratic order in fields, there are three terms at order $m^2/p^2$ (where $m$ is some soft mass), corresponding to corrections to the field self-energies:
\begin{align}
\mathcal{L}_{\rm SUSY-breaking} & = - \mathcal{C}_s \phi^* \phi - \mathcal{C}_a F^* \Box^{-1} F + i \mathcal{C}_f \chi^\dagger \sigmabar^\mu \D_\mu \Box^{-1} \chi, \label{eq:supertraceterms}
\end{align}
where the coefficients $\mathcal{C}_i$ are all $\mathcal{O}(m^2)$.  In the context of anomaly mediation, these contributions are already $\mathcal{O}(m_{3/2}^2)$, so we can neglect any further SUGRA corrections.  In particular, at this order the operator $\Box$ appearing in \Eq{eq:supertraceterms} can be thought as the d'Alembertian in flat space. 

The non-local action in \Eq{eq:supertraceterms} does not break SUSY in the limiting case $\mathcal{C}_s = \mathcal{C}_a = \mathcal{C}_f $.\footnote{Obviously, $\mathcal{C}_s$ also does not break SUSY if it arises in conjunction with a fermion mass term after an auxiliary field redefinition.  We will therefore define $\mathcal{C}_s$ to exclude such contributions.} The simple field redefinition (or the appropriately super-Weyl- and SUGRA-covariant equivalent, see \Ref{Kaplunovsky:1994fg}) 
\begin{align}
\bQ \rightarrow \bQ + \frac{\mathcal{C}}{2 \Box} \bQ \label{eq:supertracetransformation}
\end{align}
eliminates all three terms for $\mathcal{C}_i = \mathcal{C}$. Thus, a single coefficient $\mathcal{C}_i$ is not a good measure of SUSY-breaking by itself.  On the other hand, the supertrace 
\begin{align}
\mathcal{S} & = \mathcal{C}_s + \mathcal{C}_a - 2 \mathcal{C}_f, \label{eq:supertrace}
\end{align}
is invariant under the transformation of \Eq{eq:supertracetransformation} and is an unambigous measure of SUSY-breaking. \Ref{ArkaniHamed:1998kj} considered a similar supertrace over the $\mathcal{O}(m^2)$ SUSY-breaking contributions to the self-energy for the components of vector superfields.  

Of course, there is another independent combination of the $\mathcal{C}_i$ which is invariant under \Eq{eq:supertracetransformation}, which we take to be 
\begin{align}
\mathcal{T} & = \mathcal{C}_a - \mathcal{C}_f.  \label{eq:orthogonaltrace}
\end{align}
This is the unique independent choice which vanishes for tree-level SUGRA (the tachyonic scalar mass in AdS discussed in \Sec{sec:treelevel} yields vanishing $\mathcal{T}$).  A non-vanishing value of $\mathcal{T}$ is still a SUSY-breaking effect, and can be present even when the supertrace $\mathcal{S}$ vanishes.  This can arise most notably from terms like
\begin{align}
\mathcal{L} & \supset \frac{1}{\Lambda^2} \int d^4 \theta \, \frac{i}{2} \D^\dagger_{\alphadot} \bX_{\rm NL}^\dagger \sigmabar^{\mu \alphadot \alpha} \D_\alpha \bX_{\rm NL} \, \bQ^\dagger \Box^{-1} \D_\mu \bQ. \label{eq:supertraceweirdterm}
\end{align}
which yields $\mathcal{S} = 0$ but $\mathcal{T} = F_X^2/\Lambda^2$.    In the context of anomaly mediation, non-vanishing values for $\mathcal{T}$ frequently arise but they in general depend on how the theory is regulated.  In contrast, we will find that the supertrace $\mathcal{S}$ from anomaly mediation is unambiguous and irreducible, so we will mainly focus on $\mathcal{S}$ in our explicit calculations.

Analogously to \Eq{eq:supertraceterms}, there will be non-local goldstino couplings.  In the case of global flat-space SUSY, one can simply transform the terms in \Eq{eq:supertraceterms} under SUSY, with infinitesimal parameter $\epsilon = - \frac{\gold}{\sqrt{2} F_{\rm eff}}$ (see \Eq{eq:globalSUSYtransformation}), 
\begin{align}
\mathcal{L}_{\rm goldstino} & = \frac{\mathcal{G}^\mathcal{S} - \mathcal{G}^\mathcal{T} }{F_{\rm eff}} \gold \chi \phi^* +  \frac{\mathcal{G}^\mathcal{T}}{F_{\rm eff}} i \gold \sigma^\mu \D_\mu \chi^\dagger \Box^{-1} F.  \label{eq:supertracegoldstinoterms}
\end{align}
For global flat-space SUSY, $\mathcal{G}^\mathcal{S} = \mathcal{S}$ and $\mathcal{G}^{\mathcal{T}} = \mathcal{T}$.  This will not be the case, however, for AdS SUSY or for SUGRA, where there can be non-vanishing values of $\mathcal{S}$ or $\mathcal{T}$ that do not break SUSY.  Such effects will always be proportional to the inverse AdS radius $\lambda_{\rm AdS}^{-1} = m_{3/2}$.  For example, at tree level in AdS SUSY, one would use the appropriate AdS SUSY transformations (which has terms proportional to $m_{3/2}$) on the full lagrangian, which would yield $\mathcal{G}^T = \mathcal{T}$ but $\mathcal{G}^\mathcal{S} = \mathcal{S} + 2 m_{3/2}^2$.  In the following subsections, we will find these relations to be modified, but always by terms proportional to $m_{3/2}$.

\subsection{The Super-Weyl-Invariant d'Alembertian}
\label{subsec:superbox}

The operator $\superbox$ appearing in \Eq{eq:1PIscalaraction} has not been yet defined. Its definition is the last ingredient we need to computing sfermion soft masses and goldstino couplings. We will see that while $\superbox$ is generically ambiguous, our final results for the supertrace $\mathcal{S}$ and corresponding goldstino coupling $\mathcal{G}^\mathcal{S}$ are not.\footnote{This ambiguity is a reflection of an ambiguity in how to write down a SUGRA-invariant 1PI effective action, which is in addition to the ambiguity discussed in \Sec{sec:regulatorsplittings} in whether the regulators feel SUSY breaking.}

The operator $\superbox$ is a super-Weyl-invariant version of the d'Alembertian acting on scalar superfields, which reduces to $\Box$ in the limit of global flat-space SUSY.  Given a generic spinless superfield $\bU$, there are a limited number of options (neglecting fractional powers of derivatives):
\begin{align}
\superbox \bU & = \P^\dagger \P \bU + \P \P^\dagger \bU - \frac{1}{8} \frac{1}{\bC^\dagger \bC} \D^\alpha (\D^{\dagger 2} - 8 \bR) \D_\alpha \bU \nonumber \\
& + c_{1} (\bP) \P^\dagger \bU + c_{1}' (\bP^\dagger) \P \bU \nonumber + c_{2} (\bP^\dagger) \P^\dagger \bU + c_{2}' (\bP) \P \bU \nonumber\\
& + c_{3} (\P^\dagger \bP) \bU + c_{3}' (\P \bP^\dagger) \bU \nonumber + c_4 (\bP^\dagger) (\bP) \bU \nonumber \\
& + c_{5} (\bP^{\dagger 2} ) \bU +  c_{5}'  (\bP^2) \bU \nonumber + c_{6} \P^\dagger ((\bP) \bU) + c_{6}'  \P ((\bP^\dagger) \bU)  \nonumber \\
& + c_{7} \bGtilde_{\alpha \alphadot} \bC^{-1} \D^{\dagger \alphadot} \bC^{\dagger -1} \D^\alpha \bU - c_{7}' \bGtilde_{\alpha \alphadot} \bC^{\dagger -1} \D^\alpha \bC^{-1} \D^{\dagger \alphadot} \bU \nonumber \\
& + c_8 \frac{1}{\bC^\dagger \bC} \bGtilde_{\alpha \alphadot} \bGtilde^{\alpha \alphadot} \bU.    \label{eq:superboxoperators}
\end{align}
The operators and superfields $\P$, $\bP$, and $\bGtilde_{\alpha \alphadot}$ (and their hermitian conjugates) are super-Weyl covariant versions of $-\frac{1}{4} (\D^{\dagger 2} - 8 \bR)$, $2 \bR$, and $\bG_{\alpha \alphadot}$, respectively, and are defined in \App{app:superweyl}.  For matter fields $\bQ$ that are charged under a gauge group, the operators of \Eq{eq:superboxoperators} would need to be modified by appropriate insertions of $e^{\pm \bV}$.\footnote{There could also be additional possible operators proportional to the field strength $\bW_{\! \alpha}$ which would not give any contributions to self-energy corrections or goldstino couplings at the desired order.} 

Many of the terms in \Eq{eq:superboxoperators} vanish in the limit of global flat-space SUSY, so the associated coefficients $c_i$ are left completely undetermined. We could impose certain desirable properties for $\superbox$, which would lead to constraints on the $c_i$. For example, requiring that $\superbox \bU$ is chiral for chiral $\bU$ and that $\superbox$ possesses a sensible analogue of integration by parts would set $c_6 = -1$ and all other $c_i = 0$. This is the choice made in \Ref{Kaplunovsky:1994fg} (which they denote $\triangle$), though it does not satisfy $\superbox 1 = 0$.\footnote{Another obvious candidate is $\superbox = \D_a \D^a$ in the $\bC = 1$ limit (corresponding to $c_i' = c_i$, $-c_1 = c_3 = c_4/2 = c_6 = c_7 = -1/2$, $c_2 = c_5 = c_8 = 0$), though it is not chiral.}  In order to actually determine the $c_i$, one would have to explicitly take into account virtual effects to all orders in a specific regularization scheme, which is beyond the scope of this paper.  Because our final results for $\mathcal{S}$ and $\mathcal{G}^\mathcal{S}$ are independent of the $c_i$, we choose not to impose any constraint on them. 

At this point, we could use the full machinery developed in \Ref{wess1992supersymmetry} to extract the components of $\superbox \bU$.  We could then determine $\superbox^n \bU$ by recursion and find the component form of \Eq{eq:1PIscalaraction} by treating $\bZ(\superbox) \bQ$ as a Taylor expansion.\footnote{And we have.}  However, this procedure is overkill for our purposes, since we will ultimately use the trick in \Sec{subsec:SUGRAmultiplet} to find goldstino couplings once we know the dependence of the supertrace on $K_i F^i$ and $F_R$.  By super-Weyl invariance, we know our results can only depend on two parameters: 
\be
\FSW \equiv m_{3/2} + \frac{1}{3}K_i F^i \quad \text{and} \quad \frac{1}{12}\mathcal{R} \equiv m_{3/2}^2 + F_R.
\ee
Moreover, because $\mathcal{S}$ is dimension two, its only dependence on $F_R$ can be linear,\footnote{Fractional or negative powers of $m_{3/2}$ or $\mathcal{R}$ do not appear in the 1PI effective action.} so if we know the behavior of $\mathcal{S}$ for two different values of $F_R$, we can use interpolation to determine $\mathcal{S}$ for all $F_R$. Thus, it is sufficient to discuss two limiting cases where the behavior of $\superbox \bU$ simplifies.

The first limiting case is flat space but arbitrary $\vev{K_i}$.  Here, one can use the gauge choice $F_C = \FSW$ to set $M^* = 0$, and since $\mathcal{R} = 0$, one can use the global flat-space SUSY algebra to find the components of $\superbox \bU$, keeping careful track of all of the factors of $\bC$ contained therein.  In fact, one does not even need to be all that careful, by noting that 
\begin{align}
\superbox_{\rm flat} & = \frac{1}{\bC^\dagger \bC} \Box +  \left( \textrm{terms with supercovariant derivatives on } \bC, \bC^\dagger  \right). \label{eq:stupidbox}
\end{align}
There is a limited set of the possible terms in the parentheses that can contribute to physics up to $\mathcal{O}(m_{3/2}^2)$.  At $\mathcal{O}(m_{3/2})$, it can be shown explicitly that they have no effect (up to boundary terms).   At $\mathcal{O}(m_{3/2}^2)$, the effects of all such terms can be eliminated by transformations like \Eq{eq:supertracetransformation} or they take the form of \Eq{eq:supertraceweirdterm} (with $\bC$ in place of $\bX_{\rm NL}$).  In either case, they yield no contribution to the supertrace $\mathcal{S}$ of \Eq{eq:supertrace}.\footnote{Terms of the latter form do contribute to the parameter $\mathcal{T}$ defined in \Eq{eq:orthogonaltrace}, and contributions to $\mathcal{T}$ proportional to $F_R$ should still be considered SUSY-breaking.  It can be readily shown that non-zero values of $\mathcal{T}$ will only be induced by the first line of \Eq{eq:superboxoperators} or by the $c_7$ term (see \Eq{eq:Gtildedefinition}).  This $c_7$ dependence implies that the value of $\mathcal{T}$ depends on exactly how one regulates the theory.  In unbroken rigid AdS, this ambiguity does not arise; $\bG_{\alpha \alphadot} = 0$ in rigid AdS, so the term associated with $c_7$ vanishes.}  Therefore, for the purposes of finding $\mathcal{S}$ we need only consider the first term in \Eq{eq:stupidbox}, which is clearly independent of the $c_i$.  Furthermore, this is exactly the term which is already considered in the anomaly mediation literature, so the results for $\mathcal{S}$ are well-known~\cite{Randall:1998uk} (though they are usually stated as being the soft mass-squared and not the supertrace).

The second limiting case is unbroken SUSY in rigid AdS where $\vev{K_i} = 0$.  Because a flat space analysis cannot distinguish between effects proportional to $m_{3/2}^2$ (which have no associated goldstino couplings) and those proportional to $F_R$ (which do), we need a limiting case which captures terms proportional to the scalar curvature $\mathcal{R}$.  Starting with unbroken SUSY in AdS, we can luckily consider the rigid ($\MPl \rightarrow \infty$) limit without missing any physics.  The rigid AdS SUSY algebra~\cite{Adams:2011vw,deWit:1999ui,Keck:1974se,Zumino:1977av,Ivanov:1980vb} is dramatically simpler than the SUGRA algebra, corresponding to the limit $\bC = 1$, $\bR = m_{3/2}/2$, $\bG_{\alpha \alphadot} = \bW_{\alpha \beta \gamma} = 0$ \cite{Festuccia:2011ws}.  This reduces the number of independent operators in $\superbox$ to four:  
\begin{equation}
\superbox_{\rm rigid\,AdS} = \D_a \D^a - d_1 \frac{1}{4}  \D^2 - d_1' \frac{1}{4} m_{3/2} \D^{\dagger 2} + d_2 m_{3/2}^2,
\end{equation}
where the $d_i$ coefficients are related to the $c_i$ coefficients via
\begin{equation}
d_1 \equiv c_1 + c_2 + c_6, \quad  d_1'  \equiv c_1' + c_2' + c_6', \quad d_2 \equiv 2 + d_1 + d_1' + c_3 + c_3' + c_4 + c_5 + c_5' .
\end{equation}
One can then use the AdS SUSY algebra to easily extract the components of $\superbox \bU$ in AdS, find $\superbox^n \bU$ by recursion, and $\bZ(\superbox) \bQ$ by Taylor expansion.\footnote{Alternatively, one could simply work with the component form of the AdS SUSY lagrangian.  In that case, $Z(\Box)$ does not commute with SUSY transformations due to \Eq{eq:epsilonconstraint}, so one will find additional terms proportional to positive powers of $m_{3/2}$.  This approach makes it clear that the results in AdS space must be completely independent of the $c_i$, up to the transformation \Eq{eq:supertracetransformation}.}

\subsection{Soft Masses and Goldstino Couplings for Chiral Multiplets}
\label{subsec:finalanswer}

We now have all of the ingredients to determine the soft masses and goldstino couplings which follow from \Eq{eq:1PIscalaraction}.  

Applying the procedure outlined in \Sec{subsec:superbox}, we first find the behavior of $\bZ(\superbox) \bQ$ at $\mathcal{O}(m_{3/2}^2)$ in the flat space and rigid AdS limits.  Since we have argued that the final result (up to the transformation of \Eq{eq:supertracetransformation}) will depend on no parameter in \Eq{eq:superboxoperators} except for $c_7$, we will only present the answer for a choice of $c_i$ such that $\bZ(\superbox) \bQ$ is (nearly) chiral:\footnote{This choice corresponds $c_6 = -1 + c_7/2$, $c_1 = -c_7/2$, $c_3 = c_4 = -3/2$, and all other $c_i = 0$.  This is not chiral outside of AdS space, but deviations from chirality only appear in terms with gravitinos, $b_\mu$, or at $\mathcal{O}(m_{3/2}^3)$, so we neglect such terms in the following.  This choice also has the appealing feature of automatically setting $\mathcal{C}_f = 0$.}
\begin{align}
\bQ & = \phi +  \Theta \sqrt{2}\chi + \Theta^2 F, \\
\bZ(\superbox_{\rm rigid\,AdS}) & = Z(\Box) \left[\left(\phi -  \tfrac{1}{2} m_{3/2}  \gamma \Box^{-1} F + \tfrac{1}{8}m_{3/2}^2 (\gamma^2 + \gammadot - 10 \gamma) \Box^{-1} \phi \right) + \Theta \sqrt{2} \chi \right. \nonumber \\
& \qquad \qquad \left. ~+ \Theta^2 \left( F - \tfrac{1}{2}m_{3/2} \gamma \phi + \tfrac{1}{8} m_{3/2}^2 (\gamma^2 + \gammadot + 2 \gamma ) \Box^{-1} F \right) \right], \\
\bZ(\superbox_{\rm flat}) & = Z(\Box) \left[\left(\phi -  \tfrac{1}{2} \FSW  \gamma \Box^{-1} F + \tfrac{1}{8} \FSW^2 (\gamma^2 + \gammadot - 10 \gamma) \Box^{-1} \phi \right) + \Theta \sqrt{2} \chi \right. \nonumber \\
& \qquad \qquad \left. ~+ \Theta^2 \left( F - \tfrac{1}{2} \FSW \gamma \phi + \tfrac{1}{8} \FSW^2 (\gamma^2 + \gammadot + 2 \gamma ) \Box^{-1} F \right) \right] \nonumber \\
& \quad \, + Z(\Box) \tfrac{1}{2} \FSW^2 \gamma \left[  (1 - c_7) \Box^{-1} \phi + \Theta^2 (1 + c_7) \Box^{-1} F \right],
\end{align}
where the anomalous dimensions are defined as
\begin{equation}
\gamma \equiv 2 \frac{d \log Z}{d \log \Box}, \qquad \gammadot \equiv 2 \frac{d \gamma}{d \log \Box}.
\end{equation}
While $\gamma$ ($\gammadot$) is first non-zero at one-loop (two-loop) order, our results will hold to any loop order (at $\mathcal{O}(m_{3/2}^2)$).  As outlined in \Sec{subsec:superbox}, we can now find an appropriate super-Weyl invariant interpolation valid for any spacetime curvature,
\begin{align}
\bZ(\superbox) & = Z(\Box) \left(\widetilde{\phi} + \Theta \sqrt{2}  \chi + \Theta^2 \widetilde{F} \right),\\
\widetilde{\phi} & \equiv \phi -  \tfrac{1}{2} \FSW \gamma  \Box^{-1} F + \left(\tfrac{1}{8} \FSW^2 (\gamma^2 + \gammadot -  (6 + 4 c_7) \gamma )- \tfrac{1}{2} ( m_{3/2}^2 + F_R ) \gamma (1 - c_7) \right) \Box^{-1} \phi, \nonumber \\
\widetilde{F} &\equiv  F - \tfrac{1}{2 } \FSW \gamma \phi + \left( \tfrac{1}{8} \FSW^2 (\gamma^2 + \gammadot + (6 + 4 c_7) \gamma) - \tfrac{1}{2} (m_{3/2}^2 + F_R) \gamma (1 + c_7) \right) \Box^{-1} F, \nonumber 
\end{align}
remembering that $\FSW = m_{3/2}$ in flat space, and $m_{3/2}^2 + F_R$ vanishes in flat space but is $m_{3/2}^2$ for unbroken SUSY in AdS.

It is now straightforward to expand the superspace action of \Eq{eq:1PIscalaraction} (dropping factors of $Z(\Box)$ for clarity):
\begin{align}
\mathcal{L} & = \phi^* \Box \phi - i \chi^\dagger \sigmabar^\mu \D_\mu \chi +  \left|F + \frac{1}{2} (2 - \gamma) \FSW \phi \right|^2  +2  (F_R + m_{3/2}^2) \phi^* \phi \nonumber \\
&  \quad \, + \frac{1}{8} \FSW^2 (- \gamma^2 + \gammadot - (2 + 4 c_7) \gamma) \phi^*   \phi + \frac{1}{8} \FSW^2 (\gamma^2 + \gammadot + (2 + 4 c_7) \gamma) F^* \Box^{-1}  F \nonumber \\
& \quad \, - \frac{1 - c_7}{2}  ( m_{3/2}^2 + F_R )  \gamma \phi^*  \phi - \frac{1 + c_7}{2} ( m_{3/2}^2 + F_R)  \gamma F^*  \Box^{-1} F.
\end{align}
To extract the sfermion spectrum, is it helpful to perform the shift
\begin{align}
F \rightarrow F - \frac{1}{2} (2 - \gamma) \FSW \phi, \label{eq:Fshift}
\end{align}
which renders the $F$ equation of motion trivial, but induces non-zero $B$- and $A$-terms at $\mathcal{O}(m_{3/2})$ if there are superpotential terms.  Generalizing to multiple fields $\bQ^i$ with anomalous dimensions $\gamma_i$ and a superpotential
\be
\bW = \frac{1}{2} \mu_{i j} \bQ^i \bQ^j + \frac{1}{6} \lambda_{i j k} \bQ^i \bQ^j \bQ^k,
\ee
the associated scalar potential terms are
\begin{align}
V &\supset \frac{1}{2} B_{i j} \phi^i \phi^j + \frac{1}{6} A_{i j k} \phi^i \phi^j \phi^k + \rm h.c., \\
B_{i j} & = \frac{1}{2}\mu_{i j} \left(-2 + \gamma_i + \gamma_j \right) \left( m_{3/2} + \frac{1}{3} K_k F^k \right), \label{eq:anomalymediationBterm} \\
A_{i j k} & =  \frac{1}{2} \lambda_{i j k} (\gamma_i + \gamma_j + \gamma_k) \left( m_{3/2} + \frac{1}{3} K_\ell F^\ell \right), \label{eq:anomalymediationAterm} 
\end{align}
where we have expanded $\FSW = m_{3/2} + \frac{1}{3} K_i F^i$.  These are the familiar one-loop anomaly-mediated results that can be found in \Ref{Randall:1998uk,Giudice:1998xp}.

These $B$- and $A$-terms will have corresponding goldstino couplings proportional only to $K_i F^i$ but not to $m_{3/2}$.  Because the result in \Eq{eq:anomalymediationBterm} is super-Weyl invariant, we are free to choose the gauge of \Eq{eq:linearCgauge} and use the trick in \Sec{subsec:SUGRAmultiplet} to extract goldstino couplings.  For example, the $B$-term has a corresponding goldstino coupling $b_{ij}$ defined in \Eq{eq:massesandcouplings}.  Performing the shift in \Eq{eq:Thetareplacement}, we find\footnote{\label{footnote:goldstinoSWinvariance}Note that the result in \Eq{eq:bgoldstinoresult} is still invariant under the super-Weyl $F_\Sigma$ transformations.  The $K_k F^k$ factor arises by isolating the goldstino direction out of the fermion in \Eq{eq:linearCgauge}, not from $F_C$.}
\be
\label{eq:bgoldstinoresult}
b_{i j} = \frac{1}{6}\mu_{i j} \left(-2 + \gamma_i + \gamma_j \right) K_k F^k.
\ee
At $\mathcal{O}(m_{3/2})$, this goldstino coupling is independent of tuning the cosmological constant.  The difference between the $B$-term and the goldstino coupling is proportional to $m_{3/2}$
\be
B_{i j} - b_{ij} = \frac{1}{6}\mu_{i j} \left(-2 + \gamma_i + \gamma_j \right)m_{3/2},
\ee
emphasizing the role of AdS SUSY.

The key result of this paper is the sfermion supertrace $\mathcal{S}$ defined in \Eq{eq:supertrace}.  After performing the auxiliary field shift of \Eq{eq:Fshift}, we can read off the value at  $\mathcal{O}(m_{3/2}^2)$:
\begin{align}
\mathcal{S}_{i} & = - \frac{1}{4} \gammadot_{i} \left| m_{3/2} + \frac{1}{3} K_k F^k \right|^2  - (2 - \gamma_{i}) (m_{3/2}^2 + F_R). \label{eq:anomalymediationsupertrace}
\end{align}
The first term is the usual two-loop anomaly-mediated result for $\mathcal{S}$ expected from \Ref{Randall:1998uk}.   The second term is the tree-level mass splitting in AdS discussed in \Sec{sec:treelevel}, modified starting at one-loop order to include the anomalous dimension.  The fact that we have a contribution to $\mathcal{S}$ proportional to $(2 - \gamma)$ could have been anticipated, since anomaly mediation effectively tracks scale-breaking effects, and $(2 - \gamma)$ is the true scaling dimension of the operator $\bQ^\dagger \bQ$.\footnote{The same factor appeared in the auxiliary field shift of \Eq{eq:Fshift} for related reasons.}  Because $m_{3/2}^2 + F_R = \frac{1}{12}\mathcal{R}$, this second term vanishes in flat space, which is why it does not appear in the original literature.\footnote{For any negative curvature, one expects the $\gamma_{i}$ and $\gammadot_{i}$ terms to be partially cancelled off by AdS boundary effects, as in \Ref{Gripaios:2008rg}.  While we have not computed them explicitly, such boundary terms are necessary for the structure of the AdS SUSY algebra to be maintained in the unbroken limit.}   As discussed further in \App{sec:RGstability}, this whole expression is RG-stable, as it must be since it comes from a 1PI effective action.  The $\gammadot_{i}$ and $\gamma_{i}$ terms are known to be RG-stable from the general arguments in \Refs{Jack:1997eh,Jack:1999aj,Pomarol:1999ie,ArkaniHamed:1998kj}, while we argue in \App{sec:RGstability} that the tree-level result is RG-stable once one accounts for goldstino-gravitino mixing.

We can again use the trick in \Sec{subsec:SUGRAmultiplet} to extract the goldstino coupling $\mathcal{G}^{\mathcal{S}}$ defined in \Eq{eq:supertracegoldstinoterms}:\footnote{As in footnote~\ref{footnote:goldstinoSWinvariance}, the result in \Eq{eq:GSresult} is invariant under $F_\Sigma$ transformations.  $F_R$ (arising here from the gravitino equations of motion of \Eq{eq:gravitinoeoms}) does not implicitly contain $M^*M$.}  
\begin{align}
\mathcal{G}^{\mathcal{S}}_{i} & = - \frac{1}{12} \gammadot_{i} K_k F^k  \left( m_{3/2} + \frac{1}{3} K_\ell F^\ell \right) - (2 - \gamma_{i}) F_R, \label{eq:GSresult}
\end{align} 
As advertised, there are no goldstino couplings proportional to $m_{3/2}^2$.  Like $\mathcal{S}_{i}$, this associated goldstino coupling is RG-stable.  The tree-level and one-loop goldstino couplings arise because there are SUSY-preserving scalar masses in the bulk of AdS, which are then lifted by an amount proportional to the SUSY-breaking order parameter $F_R$.  For $\vev{K_i} = 0$, the two-loop anomaly-mediated masses familiar from \Ref{Randall:1998uk} have no corresponding goldstino coupling, as such masses are also present in the bulk of AdS when SUSY is unbroken.  Curiously, such two-loop goldstino couplings also vanish in the no-scale limit (where $\FSW= 0$) \cite{Lahanas:1986uc} and will be suppressed for  almost no-scale models \cite{Luty:2002hj}.  The difference between $\mathcal{S}_{i}$ and $\mathcal{G}^{\mathcal{S}}$ is
\begin{align}
\mathcal{S}_{i} - \mathcal{G}^{\mathcal{S}}_{i} &=- \frac{1}{4} \gammadot_{i} m_{3/2} \left( m_{3/2} + \frac{1}{3} K_k F^k \right)  - m_{3/2}^2 \left(2 - \gamma_{i} \right)  .
\end{align}
which is independent of the curvature $\mathcal{R}$.  As anticipated, this difference vanishes with vanishing $m_{3/2}$, as it is intimately related to SUSY-preserving anomaly mediation effects in AdS SUSY.  Whereas the second term proportional to $m_{3/2}^2$ arises purely from the structure of unbroken AdS SUSY, the first term proportional to $m_{3/2} \FSW$ is a cross term between a SUSY-preserving and a SUSY-breaking effect and vanishes in the no-scale limit.  

Results for $\mathcal{S}_{i}$ and $\mathcal{G}^{\mathcal{S}}_{i}$ are shown in \Tab{tab:summary} for various values of the curvature.  The answer is particularly striking when $\vev{K_i} = 0$ in the flat space limit with $F_R = - m_{3/2}^2$:
\begin{align}
\mathcal{S}_{i} & = - \frac{1}{4} \gammadot_{i} m_{3/2}^2, & \nonumber \\
\mathcal{G}^{\mathcal{S}}_{i} & = (2 - \gamma_{i})  m_{3/2}^2,  &(\text{flat space, $\vev{K_i} = 0$}) \\
\mathcal{S}_{i} - \mathcal{G}^{\mathcal{S}}_{i} & = - m_{3/2}^2 \left(2 - \gamma_{i} + \frac{1}{4}\gammadot_{i} \right). & \nonumber
\end{align} 
While anomaly-mediated sfermion soft mass-squareds are colloquially described as a two-loop effect, this expression makes it clear that this is an artifact of tuning the cosmological constant to zero, since anomaly mediation has important tree-level and one-loop effects on the goldstino couplings.  Indeed, the difference $\mathcal{S}_{i} - \mathcal{G}^{\mathcal{S}}_{i}$ has important effects at all orders.

For completeness, we give results for the parameter $\mathcal{T}$ defined in \Eq{eq:orthogonaltrace} and the associated goldstino coupling $\mathcal{G}^{\mathcal{T}}$: 
\begin{align}
\mathcal{T}_{i} & = - \frac{1}{8} \left(\gamma_i^2 + \gammadot_i + (2 + 4 c_7) \gamma_i \right) \left| m_{3/2} + \frac{1}{3} K_k F^k \right|^2  + \frac{1 + c_7}{2} \gamma_i (m_{3/2}^2 + F_R). \label{eq:Tresult} \\% 
\mathcal{G}^{\mathcal{T}}_{i} & = - \frac{1}{24} \left(\gamma_i^2 + \gammadot_i + (2 + 4 c_7) \gamma_i \right) K_k F^k  \left( m_{3/2} + \frac{1}{3} K_k F^k \right)  + \frac{1 + c_7}{2} \gamma_{i} F_R. \\
 \mathcal{T}_{i} - \mathcal{G}^{\mathcal{T}}_{i} & = -  \frac{1}{8}m_{3/2}\left( m_{3/2} + \frac{1}{3} K_k F^k \right)   \left(\gamma_i^2 + \gammadot_i + (2 + 4 c_7) \gamma_i \right)+  \frac{1 + c_7}{2} m_{3/2}^2 \gamma_i
\end{align} 
As expected, the difference $\mathcal{T} - \mathcal{G}^{\mathcal{T}}$ is always proportional to $m_{3/2}$, arising as it does from the structure of AdS SUSY.  However, these results are harder to interpret, since $\mathcal{T}$ has residual dependence on the parameter $c_7$ defined in \Eq{eq:superboxoperators}.  This indicates that the value of $\mathcal{T}$ depends on exactly how one regulates the theory (i.e.~on the correct choice of $\superbox$ for a given regularization scheme).  Note that if $\vev{K_i} = 0$ then $\mathcal{T} - \mathcal{G}^{\mathcal{T}}$ is independent of $c_7$.  Furthermore, in unbroken AdS SUSY ($F_R = \vev{K_i} = 0$), all $c_7$ dependence vanishes since $\bG_{\alpha \alphadot} = 0$ in rigid AdS SUSY.

\section{Conclusions}
\label{sec:discuss}

This paper completes the task originally started in \Ref{D'Eramo:2012qd} to understand anomaly mediation as being a SUSY-preserving effect in AdS space.  For the $R$-violating terms (gaugino masses, $A$-terms, and $B$-terms), anomaly mediation generates soft masses proportional to $m_{3/2}$ without corresponding goldstino couplings, making it clear that these are SUSY-preserving effects.\footnote{Strictly speaking, we have not carried out the calculation of gaugino masses beyond one-loop order.  We sketch how to do this in \App{sec:1PIgaugino}.}  For the sfermion soft mass-squareds, the situation is far more interesting, since there are SUSY-preserving effects proportional to $m_{3/2}^2$ and SUSY-breaking effects proportional to $F_R$, but these two effects are difficult to disentangle because $F_R$ happens to equal $-m_{3/2}^2$ after tuning for flat space.  Having successfully isolated these two effects, we see that the familiar two-loop anomaly-mediated sfermion soft mass-squareds are accompanied by tree-level and one-loop goldstino couplings, and all three terms are needed to preserve the underlying  AdS SUSY structure.

Along the way, we have learned a number of lessons about AdS SUSY and SUGRA.  First, the peculiar behavior of anomaly mediation is already evident at tree-level, and the irreducible goldstino coupling in \Eq{eq:irreduciblegoldstino} offers strong evidence that AdS SUSY (and not flat space SUSY) is the correct underlying symmetry structure for SUGRA theories.  Second, to incorporate quantum effects, one has to work with a regulated SUGRA action.  Unfortunately, it is impossible to write down a Wilsonian action that captures the full effects of anomaly mediation at tree-level, since there are important effects of the regulator fields at loop-level.  Instead, we used a 1PI effective action to make super-Weyl invariance manifest, countering the (gauge-dependent) claims in \Refs{deAlwis:2008aq,deAlwis:2012gr} (and implicit in \Ref{Kaplunovsky:1994fg}) about the non-existence of anomaly mediation.  Third, even with a SUSY-preserving, super-Weyl-invariant 1PI effective action in hand, there is residual ambiguity starting at $\mathcal{O}(m_{3/2}^2)$ in how to write down a SUGRA-invariant theory.  Luckily, the supertrace $\mathcal{S}$ is unambiguous, yielding the same soft mass-squareds known in the literature.

This paper has focused on formal aspects of anomaly mediation, and therefore has not addressed a number of important phenomenological questions.  First, anomaly mediation was motivated in part by the possibility of sequestering, and one would like to know whether the sequestered limit is physically obtainable without fine-tuning.  To that end, it would be useful to know whether the irreducible goldstino coupling in \Eq{eq:irreduciblegoldstino} is indeed an attractive IR fixed point, as one would expect in conformally sequestered theories.  Second, we have used goldstino couplings as a probe of which effect preserve SUSY and which effects break SUSY.  Ideally, one would want to find an experimental context where these goldstino couplings could be measured, since this would give an experimental handle on the underlying AdS curvature.  Measuring such a coupling to two-loop precision would even probe the value of $\FSW$, though the physical significance of that dependence is not clear to us.  Third, in addition to the supertrace $\mathcal{S}$, we identified the independent trace $\mathcal{T}$ which is perhaps known to SUSY aficionados but is unfamiliar to us.  Even in global flat space SUSY, it would be helpful to know what effects a non-zero value of $\mathcal{T}$ can have on phenomenology.  Finally, the big question facing particle physics in 2013 is whether (weak scale) SUSY is in fact realized in nature.  We of course have no insight into this broader question, but we can say that if (AdS) SUSY and SUGRA do exist, then anomaly mediation will yield irreducible physical effects proportional to $m_{3/2}$.

\begin{acknowledgments}
We benefited from conversations with Allan Adams, Daniel Freedman, Markus Luty, Yasunori Nomura, Riccardo Rattazzi, and Raman Sundrum.  We thank Senarath de Alwis for vigorous discussions on these issues.  J.T. and Z.T. are supported by the U.S. Department of Energy (DOE) under cooperative research agreement DE-FG02-05ER-41360.  J.T. is also supported by the DOE under the Early Career research program DE-FG02-11ER-41741.  F.D. is supported by the Miller Institute for Basic Research in Science.
\end{acknowledgments}

\appendix

\section{Goldstino Couplings from the Conformal Compensator}
\label{app:compensator}

In this appendix, we provide a third derivation of the goldstino couplings in \Eq{eq:goldstinocouplings}, working in the conformal compensator formalism of SUGRA to connection to our previous analysis in \Ref{D'Eramo:2012qd}.\footnote{For details on the conformal compensator formalism see \Refs{Siegel:1978mj,Gates:1983nr,Kugo:1982cu}.  This formalism is reviewed in \Ref{Cheung:2011jp} using two-component fermion notation.}  Here, the extra gauge redundancies of conformal SUGRA are gauge fixed to recover minimal SUGRA~\cite{Kaku:1977pa,Kaku:1978ea,Kaku:1978nz,Townsend:1979ki} via a conformal compensator $\bPhi$, a chiral field with conformal weight $1$.  We can use $\bPhi$ to build a superconformally invariant action at tree-level (dropping Yang-Mills terms for convenience)
\begin{equation}
\label{eq:superconformallagrangian}
\mathcal{L} = \int d^4 \theta \,  \bPhi^\dagger \bPhi \, \bOmega + \int d^2 \theta \, \bPhi^3 \bW + \textrm{h.c.} + \ldots, \qquad  \bOmega \equiv -3 e^{-\bK /3} \ .
\end{equation}
Here, we use global superspace variables to express only the matter parts of the action, and the ellipsis ($\ldots$) represents the action for the gravity multiplet as well as couplings of the matter fields to the gravity multiplet (see, e.g.,  \Refs{Kugo:1982cu,Cheung:2011jp}).

The gauge choice for $\bPhi$ proposed by Kugo and Uehara~\cite{Kugo:1982mr} allows us to use the ``global superspace'' terms of \Eq{eq:superconformallagrangian} to find the pertinent features of supergravity, including scalar masses and goldstino couplings in curved space, without having to worry about supergravity effects from the terms in the ellipsis.\footnote{An alternative gauge fixing was proposed in \Ref{Cheung:2011jp}, but it is only valid in flat space. Given this limitation, it would obfuscate the derivation of the sfermion spectrum in curved space.} This gauge is
\be
\bPhi  = e^{K/6 - i / 3 \Arg W} \left\{ 1, \frac{1}{3} K_i \chi^i , F_\Phi \right\} ,
\label{eq:KUgauge}
\ee
where the field $F_\Phi$ is an auxiliary complex degree of freedom, corresponding to the complex auxiliary field $M$ of supergravity.  Unlike in the super-Weyl formalism, $F_\Phi$ is not a gauge degree of freedom.

The most general K\"ahler and superpotential for unbroken SUGRA in AdS (i.e.~$\vev{W_i} = \vev{K_i} = 0$) is\footnote{For simplicity, we assume none of the visible-sector fields are singlets.  The physics does not appreciably change if there are singlets, as long as there is no SUSY breaking in the visible sector.}
\begin{align}
\bOmega & = \bQ^{\dagger \i} \bQ^i + \frac{1}{2} \vev{\Omega_{ij}} \bQ^i \bQ^j + \textrm{h.c.} + \ldots, \\
\bW & = m_{3/2} + \frac{1}{2} \vev{W_{ij}} \bQ^i \bQ^j + \ldots,
\end{align}
where the ellipses represent higher-order terms. Inserting these expression into \Eq{eq:superconformallagrangian} and rescaling the fields $\bQ^i \rightarrow \bQ^i / \bPhi$, we can solve the $F_\Phi$ equation of motion to find $F_\Phi  = m_{3/2} + \ldots$.  The extra terms are suppressed by at least two powers of $\MPl$, and thus irrelevant for our purposes. It is then simple to read off the cosmological constant, as well as the fermion and scalar mass matrices:
\begin{align}
\vev{V} & = - 3 m_{3/2}^2 \MPl^2, \label{eq:AdSCC} \\
M_{i j} & = \vev{W_{i j}}+ m_{3/2} \vev{\Omega_{i j}} , \\
m^2_{i \j} & = M_{i k} M^k{}_\j - 2 m_{3/2}^2 \delta_{i \j}, \label{eq:AdSsoftmasses}\\
B_{i j} & = - m_{3/2} \vev{W_{i j}} + m_{3/2}^2 \vev{\Omega_{i j}} - 2 m_{3/2}^2 \vev{\Omega_{i j}} = - m_{3/2} M_{i j}. \label{eq:AdSBterms}
\end{align}
Thus, we recover the universal tachyonic soft mass-squared in \Eq{eq:unbrokenSUSYspectrum} for scalars in unbroken AdS SUGRA, as well as $B$-terms proportional to the fermion mass matrix.  

SUSY breaking effects then lift AdS space up to flat space. We represent the source of SUSY breaking in the hidden sector by a non-linear goldstino multiplet~\cite{Rocek:1978nb,Lindstrom:1979kq,Komargodski:2009rz,Cheung:2010mc,Cheung:2011jq}
\be
\bX_{\NL} = F_X \left(\theta + \frac{1}{\sqrt{2} F_X}  \gold \right)^2,
\ee
where $\gold$ is the goldstino.  Because of the constraint $\bX_{\NL}^2 = 0$, the K\"ahler potential and superpotential terms involving the non-linear field $\bX_{\NL}$ are strongly constrained
\begin{align}
\bOmega & \supset -3 + \vev{\Omega_X} \bX_{\NL} + \vev{\Omega_\X} \bX_{\NL}^\dagger + \vev{\Omega_{X \X}}  \bX_{\NL}^\dagger \bX_{\NL}, \\
\bW & \supset m_{3/2} + \vev{W_X} \bX_{\NL}  \label{eq:WwithX}.
\end{align}
The coefficients $\vev{\Omega_X}$ and $\vev{W_X}$ can be made real by using our freedom to rotate $\bX_{\NL}$ and perform K\"ahler transformations. A canonically-normalized goldstino (i.e.~$\bK \supset \bX_{\NL}^\dagger \bX_{\NL}$) enforces the condition $\vev{\Omega_{X \X}} = 1 - \frac{1}{3} \vev{\Omega_X}^2$. Upon rescaling the non-linear field $\bX_{\NL} \rightarrow \bX_{\NL} / \bPhi$ and integrating out auxiliary fields, we find from \Eq{eq:superconformallagrangian}:
\begin{align}
\vev{F_X} & = - \vev{W_X - m_{3/2} \Omega_X}, \\
\vev{F_\Phi} & = m_{3/2} + \frac{1}{3} \vev{\Omega_X F^X}, \\
\vev{V} & = \vev{F_X^2} - 3 m_{3/2}^2.
\end{align}
The amount of SUSY breaking to achieve flat space is thus $\vev{F_X} = \sqrt{3} m_{3/2}$.  We also have a canonically-normalized goldstino with mass $2 m_{3/2}$ \cite{Cheung:2010mc,Cheung:2011jq}.

The K\"ahler potential and superpotential will also include direct couplings between visible matter fields and the SUSY breaking sector. For simplicity, we start our study of goldstino couplings for massless visible sector fermions (e.g.~$\bQ^i \bQ^j$ is never a singlet under any of the gauge symmetries in the theory).  In this simple case the operators we can add are
\begin{align}
\label{eq:PhiMatterXcouplings}
\bOmega & \supset \vev{\Omega_{i \j X \X}}   \bQ^{\dagger \j} \bQ^i \bX_{\NL}^\dagger \bX_{\NL}  , \\
\bW & \supset \frac{1}{6} \vev{W_{i j k}}  \bQ^i \bQ^j \bQ^k +  \frac{1}{6} \vev{W_{i j k X}}  \bQ^i \bQ^j \bQ^k \bX_{\NL} ,
\end{align}
where we have eliminated any possible $\bQ^{\dagger \j} \bQ^i \bX_{\NL}$ terms by using our freedom to perform a transformation $\bQ^i \rightarrow \bQ^i + n^i{}_j \bQ^j \bX_{\NL}$~\cite{D'Eramo:2012qd}. The scalar masses and $A$-terms can be easily read off from \Eq{eq:superconformallagrangian}:
\begin{align}
m^2_{i \j} & = - 3 m_{3/2}^2 \vev{\Omega_{i \j X \X}}, \\
A_{i j k} & = \sqrt{3} m_{3/2} \vev{W_{i j k X}}.
\end{align}

The terms in \Eq{eq:PhiMatterXcouplings} also yield goldstino couplings to visible sector fields from the fermionic component of $\bX_{\NL}$; namely $a_{i \j} \supset m^2_{i \j}$.  Less obvious is that there are additional goldstino couplings coming from $\bPhi$.  In the gauge from \Eq{eq:KUgauge}, the fermionic component of $\bPhi$ contains visible sector fermions (coupled to its conjugate scalar):  
\be
\frac{1}{3} K_i \chi^i =  \frac{1}{3} \vev{\Omega_X} \gold{} + \frac{1}{3} \phi^{* \i} \chi^i + \ldots
\ee  
This means that the $\vev{W_X} \bX_{\NL}$ term in the superpotential of \Eq{eq:WwithX} (multiplied by $\bPhi^2$ after rescaling) gives an additional coupling $(2 m_{3/2}^2 / F_X) \, K_i \chi^i \gold$ (i.e.~the universal goldstino couplings from \Eq{eq:irreduciblegoldstino}).  The full goldstino coupling reads
\be
a_{i \j} = m^2_{i \j} + 2 m_{3/2}^2 \delta_{i \j} \ ,
\ee
in agreement with \Eq{eq:goldstinocouplings} in the $M_{i j} = 0$ limit.

Finally, we consider superpotential and Giudice-Masiero mass terms for the fermions. This introduces a plethora of new possible terms:
\begin{align}
\bOmega & \supset  \frac{1}{2} \bQ^i \bQ^j \left[ \vev{\Omega_{i j}} + \vev{\Omega_{i j X}} \bX_{\NL} +  \vev{\Omega_{i j \X}}  \bX_{\NL}^\dagger + 
\vev{\Omega_{i j X \X}} \bX_{\NL}^\dagger \bX_{\NL}  \right]  , \\
\bW & \supset  \frac{1}{2} \vev{W_{i j}} \bQ^i \bQ^j + \frac{1}{2} \vev{W_{i j X}} \bQ^i \bQ^j \bX_{\NL} .
\end{align}
Fermion masses and $B$-terms can be easily extracted from this lagrangian.  Goldstino couplings are more difficult to read off.  As already mentioned, the goldstino lives both in $\bPhi$ and $\bX_{\NL}$, but in addition, the K\"ahler potential cubic terms $\bQ^i \bQ^j \bPhi^\dagger$ and $\bQ^i \bQ^j \bX_{\NL}^\dagger$ contain derivative interactions with the goldstino.  After using the equation of motion for the goldstino of mass $2 m_{3/2}$ 
\be
\phi^j \chi^i (-i \sigma^\mu \partial_\mu \gold^\dagger) \rightarrow 2 m_{3/2} \phi^j \chi^i \gold,
\ee
these yield Yukawa interactions between matter fields and the goldstino.\footnote{The problematic cubic term $\bQ^i \bQ^j \bPhi^\dagger$ could have been eliminated by a redefinition of $\bPhi$, or equivalently choosing a different gauge fixing than the one in \Eq{eq:KUgauge}.  The $\bQ^i \bQ^j \bX^{\dagger}_{\rm NL}$ term, however, cannot be eliminated by any redefinition that preserves $\bX_{\rm NL}^2 = 0$. }  The resulting goldstino couplings are exactly those of \Eq{eq:goldstinocouplings}.

\section{Renormalization Group Invariance of Irreducible Goldstino Couplings}
\label{sec:RGstability}

In \Sec{sec:treelevel}, we found a universal tree-level goldstino coupling to matter scalars and fermions proportional in $m_{3/2}^2$.  In \Sec{sec:twoloop}, we expanded this result to all loop orders, finding further couplings by carefully analyzing the SUGRA- and super-Weyl invariant 1PI effective action:
\be
\mathcal{G}^{\mathcal{S}}_i  =  2 m_{3/2}^2 - \gamma_i m_{3/2}^2 - \frac{1}{12} \gammadot_i K_j F^j  \left( m_{3/2} + \frac{1}{3} K_j F^j \right)  \qquad (\text{flat space}).\label{eq:appendixgoldstinosupertrace}
\ee
Since these results follow from a 1PI action, they have incorporated all quantum corrections and are thus completely RG stable---that is, their coefficients solve their own RG equations.  For the terms proportional to $\gamma_i$ and $\gammadot_i$, it has long been known in the literature \cite{Jack:1997eh,Jack:1999aj,Pomarol:1999ie,ArkaniHamed:1998kj} that mass terms of such a form are RG stable.  This is true for the $\gamma_i$ term by itself, and is true for the $\gammadot_i$ term given corresponding $A$ terms in the form of \Eq{eq:anomalymediationAterm}.  The same logic for soft terms can be trivially extended to goldstino couplings, which makes it clear that the goldstino couplings proportional to $\gamma_i$ and $\gammadot_i$ above are also RG stable.\footnote{This logic is less clearly applicable for the $\gammadot_i m_{3/2} K_j F^j$ crossterm, as the goldstino coupling corresponding to the $A$-terms of \Eq{eq:anomalymediationAterm} is not expected to depend on $m_{3/2}$.  Nevertheless, the logic still holds.}

However, the tree-level term, proportional to a constant, is not so clearly RG stable.   Naively, one would expect it to receive quantum corrections starting at one loop (separate from the term proportional to $\gamma$ in \Eq{eq:appendixgoldstinosupertrace}), just as a constant scalar mass would.  This puzzle is resolved by remembering that the goldstino and gravitino mix in SUGRA, so quantum corrections to gravitino couplings feed into quantum corrections to goldstino couplings, making the tree-level goldstino coupling in \Eq{eq:appendixgoldstinosupertrace} RG stable.

\begin{figure}
  \centering

  \includegraphics[width=150pt]{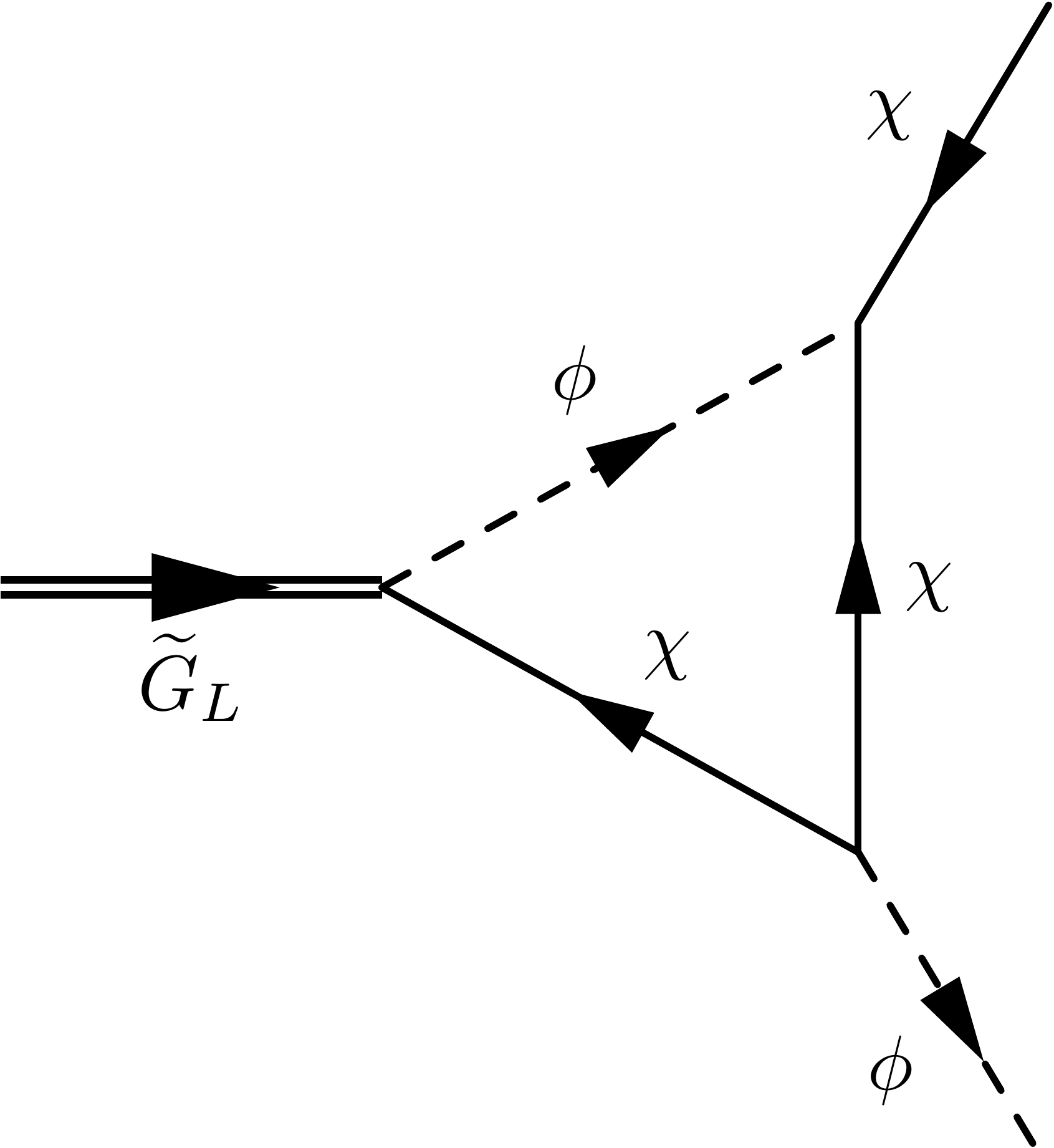}
    \caption{One-loop diagram that renormalizes the goldstino coupling to visible-sector scalars and fermions in the Wess-Zumino theory from \Eq{eq:WZsimpletheory}.  The diagram has the same logarithmic divergence in both global SUSY and SUGRA, and would seem to renormalize the tree-level goldstino coupling $\mathcal{G}^{\mathcal{S}}_i  \supset  2 m_{3/2}^2$.}
    \label{fig:goldstinotriangle}
\end{figure}

%\begin{fmffile}{goldstinotriangle}
%\begin{figure}
%  \centering

%    \begin{fmfgraph*}(150,150)
   %   \fmfleft{i1}
     % \fmfright{o1,o2}
     % \fmf{dbl_plain_arrow,label=$\widetilde{G}_L$}{i1,v1}
     % \fmf{fermion,tension=0.4,label=$\chi$}{v3,v1}
    % \fmf{scalar,tension=0.4,label=$\phi$}{v1,v2}
     % \fmf{fermion,tension=0.4,label=$\chi$}{v3,v2}
     % \fmf{scalar,label=$\phi$}{v3,o1}
     % \fmf{fermion,label=$\chi$}{o2,v2}
   % \end{fmfgraph*}
   % \caption{One-loop diagram that renormalizes the goldstino coupling to visible-sector scalars and fermions in the Wess-Zumino theory from \Eq{eq:WZsimpletheory}.  The diagram has the same logarithmic divergence in both global SUSY and SUGRA, and would seem to renormalize the tree-level goldstino coupling $\mathcal{G}^{\mathcal{S}}_i  \supset  2 m_{3/2}^2$.  \jdt%{The arXiv barfs with feynmf.  Please turn into a PDF file.  If you can't do it, I can do with with LaTeXiT.}}
  %  \label{fig:goldstinotriangle}
%\end{figure}
%
%\end{fmffile}

For clarity, we give an example of how this occurs in one concrete model: a sequestered theory (in the sense of \Eq{eq:factorized_form}) in flat space with $\vev{K_i} = 0$ and a Wess-Zumino visible sector:
\be
\label{eq:WZsimpletheory}
\bW_{\rm vis}  = \frac{1}{6} \lambda \bQ^3,
\ee
with $\bQ = \{\phi, \chi, F \}$. The goldstino coupling seems to receive a correction from the logarithmically divergent diagram in \Fig{fig:goldstinotriangle}. Using a Pauli-Villars regulator, the divergent part of this diagram is 
\be
i \mathcal{M}_1 = i \frac{2 m_{3/2}^2}{F_{\rm eff}} x_{\gold} y_\chi \left( - \frac{\lambda^2}{(4 \pi)^2} \log \Lambda^2 \right) + \ldots \label{eq:triangledivergence1},
\ee
with $x_{\gold}$ and $y_\chi$ the external wave function spinors for the goldstino and the visible-sector fermion, respectively.\footnote{We use the methods of \Ref{Dreiner:2008tw} for calculations here, but keep the sign and sigma matrix conventions of \Ref{wess1992supersymmetry}.}
The presence of such a divergence would be fine if it could be completely absorbed by the wave-function renormalization of the visible sector fields.  However, we know that it cannot be absorbed in the global SUSY case, which features the exact same diagram (up to a soft scalar mass that does not affect its divergent part).  Explicitly, one can see this by noting that the divergent one-loop contribution to $Z$ is
\begin{align}
Z & = \frac{1}{2} \frac{\lambda^2}{(4 \pi)^2} \log \Lambda^2 + \ldots.  \label{eq:Zdivergence}
\end{align}
This differs by a factor of $-2$ from what would be needed to have the entire divergence in \Eq{eq:triangledivergence1} explained by wave function renormalization.  Thus, one would seem to find that the $\mathcal{G}^{\mathcal{S}}_i  \supset  2 m_{3/2}^2$ goldstino coupling runs at one-loop order, in conflict with the claims that $\mathcal{G}^{\mathcal{S}}_i$ arises from a valid 1PI effective action.

What we have not accounted for, however, is the mixing between the gravitino and the goldstino in SUGRA.  Recall that the equation of motion of the gravitino in flat space is  
\be
\sigma^{\mu \nu} \D_\mu \psi_\nu = \sqrt{\frac{3}{2}} m_{3/2} \gold + \frac{3}{4} i m_{3/2} \sigma^\mu \psi^\dagger_\mu , \label{eq:gravitinoeomappendix}
\ee
so diagrams with an external gravitino may yield corrections to the goldstino coupling after using this equation of motion (or making an appropriate field redefinition).\footnote{One can of course pick a gauge for the Rarita-Schwinger gravitino field which removes the the quadratic mixing and changes this equation of motion.  As in the text, we will only pick a gauge for the gravitino-goldstino system after computing quantum corrections to all orders in visible-sector couplings.  This does not pose a problem as we never have to consider gravitinos or goldstinos (whose couplings are suppressed by $\MPl^{-1}$) as internal legs when computing such quantum corrections.}  Effectively, by trading away couplings proportional to the left-hand side of \Eq{eq:gravitinoeomappendix}, we are making sure that we are still in Einstein frame at one-loop order.

\begin{figure}
  \centering
  	\includegraphics[width=150pt]{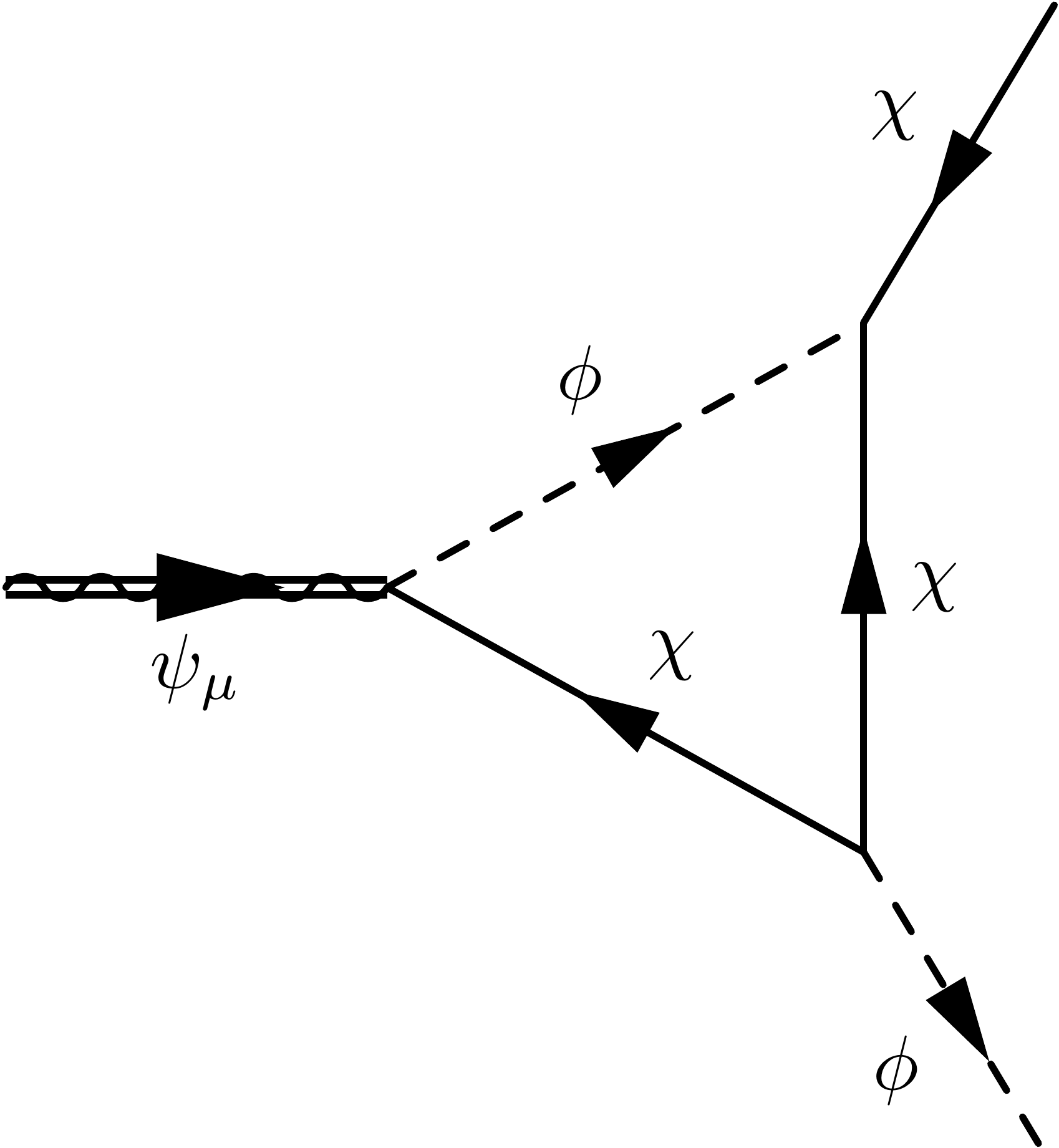} $\qquad$
	\includegraphics[width=150pt]{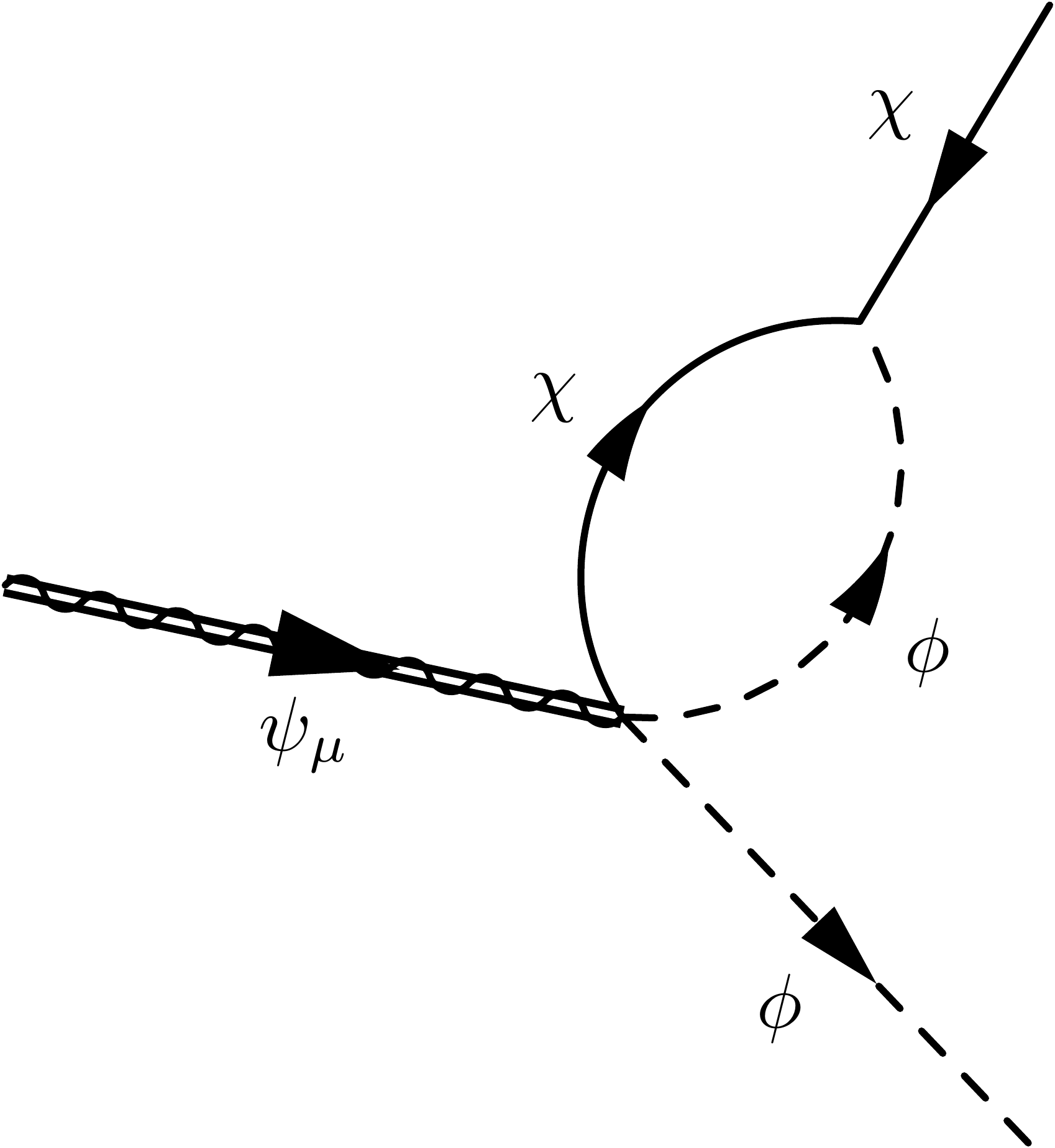}
	
    \caption{These two diagrams yield logarithmically divergent corrections to the goldstino coupling after using the equation of motion in \Eq{eq:gravitinoeomappendix} for the gravitino.  When combined with the diagram in \Eq{fig:goldstinotriangle}, the goldstino coupling $\mathcal{G}^{\mathcal{S}}_i  \supset  2 m_{3/2}^2$ is RG stable.}
    \label{fig:gravitinotriangle}

\end{figure}

%\begin{fmffile}{gravitinotriangle}

%\begin{figure}
 % \centering
   % \begin{fmfgraph*}(150,150)
   %   \fmfleft{i1}
    %  \fmfright{o1,o2}
     % \fmf{dbl_plain_arrow,tension=0.5,label=$\psi_\mu$}{i1,v1}
     % \fmf{photon,tension=0.5}{i1,v1}
     % \fmf{fermion,tension=0.4,label=$\chi$}{v3,v1}
    % \fmf{scalar,tension=0.4,label=$\phi$}{v1,v2}
    %  \fmf{fermion,tension=0.4,label=$\chi$}{v3,v2}
     % \fmf{scalar,label=$\phi$}{v3,o1}
    %  \fmf{fermion,label=$\chi$}{o2,v2}
   % \end{fmfgraph*}
   % \begin{fmfgraph*}(150,150)
    %  \fmfleft{i1}
    %  \fmfright{o1,o2}
    %  \fmf{dbl_plain_arrow,tension=0.5,label=$\psi_\mu$}{i1,v1}
    %  \fmf{photon,tension=0.5}{i1,v1}
     % \fmf{fermion,left=0.625,tension=0.4,label=$\chi$}{v1,v2}
    % \fmf{scalar,right=0.625,tension=0.4,label=$\phi$}{v1,v2}
    %  \fmf{scalar,label=$\phi$}{v1,o1}
    %  \fmf{fermion,label=$\chi$}{o2,v2}
   % \end{fmfgraph*}
  %  \caption{These two diagrams yield logarithmically divergent corrections to the goldstino coupling after using the equation of motion in \Eq{eq:gravitinoeomappendix} for the gravitino.  When combined with the diagram in \Eq{fig:goldstinotriangle}, the goldstino coupling $\mathcal{G}^{\mathcal{S}}_i  \supset  2 m_{3/2}^2$ is RG stable.}
 %   \label{fig:gravitinotriangle}

%\end{figure}

%\end{fmffile}

Using $G$ defined in \Eq{eq:Gdef}, the gravitino couples to visible-sector fields as \cite{wess1992supersymmetry} 
\begin{align}
\mathcal{L} & = - \frac{1}{\sqrt{2} \MPl} g_{i \j} \partial_\nu \phi^{* \j} \chi^i \sigma^\mu \sigmabar^\nu \psi_\mu - e^{\frac{G}{2 \MPl^2}} \frac{i}{\sqrt{2}} G_i \chi^i \sigma^\mu \psi^\dagger_\mu + \rm h.c. \\
& = - \sqrt{\frac{3}{2}} \frac{m_{3/2}}{F} \psi_\mu \sigma^\nu \sigmabar^\mu \chi \partial_\nu \phi^* + \frac{1}{2} i \lambda \sqrt{\frac{3}{2}} \frac{m_{3/2}}{F} \psi_\mu \sigma^\mu \chi^\dagger \phi^{* 2} + \ldots  +  \rm h.c.,
\end{align}
where in the second line we have specialized to the theory in \Eq{eq:WZsimpletheory}.  The two diagrams featuring an external gravitino that can give contributions proportional to the left-hand side of \Eq{eq:gravitinoeomappendix} are shown in \Fig{fig:gravitinotriangle}.  Each of these diagrams is logarithmically divergent,\footnote{In fact, they are linearly divergent, but any ensuing subtleties will only affect the finite pieces, not the logarithmically divergent ones.} and they give equal corrections to the goldstino coupling.  Combining these with \Eq{eq:triangledivergence1}, we find
\be
i \mathcal{M}_{\rm total} = i \frac{2 m_{3/2}^2}{F_{\rm eff}} x_{\gold} y_\chi \left( \frac{1}{2} \frac{\lambda^2}{(4 \pi)^2} \log \Lambda^2 \right) + \ldots \label{eq:triangledivergenceFixed}.
\ee
Comparing this to \Eq{eq:Zdivergence}, we see this is precisely the logarithmic divergence that can be completely absorbed by the wave function renormalization of the visible-sector fields.  At the one-loop level in this model, we confirm that the tree-level goldstino coupling does not run, as we knew had to be the case from our 1PI analysis in \Sec{sec:twoloop}.

\section{Super-Weyl Transformations}
\label{app:superweyl}

 Super-Weyl transformations are the most general transformations that leave the torsion and chirality constraints of SUGRA unchanged.  They may be completely parameterized by a chiral superfield $\bSigma$ and its conjugate anti-chiral superfield $\bSigma^\dagger$ \cite{Howe:1978km,wess1992supersymmetry}.  The super-Weyl transformations act infinitesimally on the gravity multiplet as \cite{Howe:1978km,wess1992supersymmetry,Kaplunovsky:1994fg} 
\begin{align}
\delta E_M{}^a & = (\bSigma + \bSigma^\dagger) E_M{}^a, & \delta E_M{}^\alpha & = (2 \bSigma^\dagger - \bSigma) E_M{}^\alpha - \frac{i}{2} E_M{}^a (\D^{\dagger}_\alphadot \bSigma^\dagger \sigmabar^{\alphadot \alpha}_a), \nn \\
\delta \D_\alpha & = (\bSigma - 2 \bSigma^\dagger) \D_\alpha -2 (\D^\beta \bSigma) L_{\alpha \beta}, & \delta \D^\dagger_{\alphadot} & = (\bSigma^\dagger - 2 \bSigma) \D^\dagger_{\alphadot} - 2 (\D^{\dagger \betadot} \bSigma^\dagger) L_{\alphadot \betadot},  \nn \\
\delta \bE & = 2 (\bSigma + \bSigma^\dagger) \bE, & \delta (2 \bEpsilon) & = 6 \bSigma (2 \bEpsilon) + \ldots, \nn \\
\delta \bR & = 2 ( \bSigma^\dagger - 2 \bSigma) \bR - \frac{1}{4} \D^{\dagger 2} \bSigma^\dagger, & \delta \bG_{\alpha \alphadot} & = - (\bSigma + \bSigma^\dagger) \bG_{\alpha \alphadot} + i \D_{\alpha \alphadot} (\bSigma^\dagger - \bSigma), \nn \\
\delta \bW_{\alpha \beta \gamma} & = - 3 \bSigma \bW_{\alpha \beta \gamma}, \end{align}
where $a$ is a local Lorentz spacetime index, $L_{\alpha \beta}$ are the Lorentz generators acting on spinors, $\bE$ is the determinant of the supersymmetric vielbein, $2 \bEpsilon$ is the corresponding chiral density, $\bR$ is the chiral curvature superfield, and $\bG_{\alpha \alphadot}$ is the real superfield having the vector auxiliary field of supergravity $b_\mu$ as its lowest component.  The ellipsis in the transformation of the chiral vielbein are omitted terms irrelevant for the construction of a super-Weyl invariant action.  The transformation of $\D_a$ is too complicated to include here, but $\D_a$ may always be expressed as some composition of the above objects.  For example, when acting on a Lorentz scalar superfield $\bU$,
\begin{align}
\D_a \bU & = - \frac{1}{4} i \sigmabar_{a}^{\alphadot \alpha} \{ \D^{\dagger}_{\alphadot}, \D_\alpha \} \bU.
\end{align}

Chiral superfields $\bQ$ and vector superfields $\bV$ transform as \cite{wess1992supersymmetry} 
\begin{align}
\delta \bQ & = w \bSigma \bQ, & \delta \bV & =  w' (\bSigma + \bSigma^\dagger) \bV,
\end{align}
where $w$ and $w'$ are the Weyl weights of their respective superfield; for ordinary matter or gauge superfields, these weights vanish.  Note that the higher components of matter superfields still transform, due to the non-trivial transformation of the $\D_\alpha$ used to project them out.  For a vector superfield of weight 0, the superfield
\begin{align}
\bW_{\! \alpha} & \equiv - \frac{1}{4} (\D^{\dagger 2} - 8 \bR) \D_\alpha \bV \label{eq:Walphadefinition}
\end{align}
transforms as a chiral superfield of Weyl weight $-3$.

The SUGRA action of \Ref{wess1992supersymmetry} can be made super-Weyl invariant by including a super-Weyl compensator $\bC$ of Weyl weight $-2$.  The tree-level lagrangian then reads
\begin{align}
\mathcal{L} & = \int d^4 \Theta \, \bE \, \bC^\dagger \bC \, (-3 e^{-\bK/3}) + \int d^2 \Theta \, 2 \bEpsilon \, \bC^3 \bW + \frac{1}{4} \int d^2 \Theta \, 2 \bEpsilon \, \bW^\alpha \bW_{\! \alpha} + {\rm h.c.}  \label{eq:superweyllagrangianAppendix}
\end{align}
The super-Weyl compensator can also be used to build versions of $\bR$ and $\bG_{\alpha \alphadot}$ that transform homogeneously under super-Weyl transformations:  
\begin{align}
\bP & \equiv - \frac{1}{4} \frac{1}{\bC^2} (\D^{\dagger 2} - 8 \bR) \bC^\dagger, \label{eq:Pdefinition} \\
\bP^\dagger & \equiv - \frac{1}{4} \frac{1}{\bC^{\dagger 2}} (\D^2 - 8 \bR) \bC, \\
\delta \bP & = \delta \bP^\dagger  = 0, \\
\widetilde{\bG}_{\alpha \alphadot} & \equiv \bG_{\alpha \alphadot} - \frac{1}{4 \bC^\dagger} \D_\alpha \D^\dagger_{\alphadot} \bC^\dagger + \frac{1}{4 \bC} \D^\dagger_{\alphadot} \D_\alpha \bC + \frac{1}{4 \bC^\dagger \bC} (\D_\alpha \bC) (\D^\dagger_{\alphadot} \bC^\dagger), \label{eq:Gtildedefinition} \\
\delta \bGtilde_{\alpha \alphadot} & = - (\bSigma + \bSigma^\dagger) \bGtilde_{\alpha \alphadot}.
\end{align}
These objects also obey appropriately-modified versions of the Bianchi identities:
\begin{align}
\D^\dagger_{\alphadot} \bP & = 0, &  \D_\alpha \bP^\dagger & = 0, \\
\D^\alpha ( \bC \bGtilde_{\alpha \alphadot} ) & = \frac{1}{2} \bC^{\dagger 2} \D^\dagger_{\alphadot} \bP^\dagger, & \D^{\dagger \alphadot} (\bC^\dagger \bGtilde_{\alpha \alphadot} ) & = \frac{1}{2} \bC^2 \D_\alpha \bP.
\end{align}
The superfield $\bP$ ($\bP^\dagger$) can also be serve as an operator, which we denote by the non-boldface $\P$ ($\P^\dagger$).  When acting on a super-Weyl invariant spinless superfield, $\P$ ($\P^\dagger$) returns a super-Weyl invariant (anti-)chiral superfield \cite{Kaplunovsky:1994fg}.  The operator $\P$ ($\P^\dagger$) thus acts as an (anti-)chiral projector.

\section{1PI Gaugino Masses}
\label{sec:1PIgaugino}

In \Eq{eq:chiral1PI}, we used a 1PI effective action for the gauge multiplet built as an integral over chiral superspace.  This is sufficient for extracting one-loop results, but in a general renormalization scheme, the 1PI action must instead be written as an integral of a non-local quantity over all of superspace.  For the familiar case of global SUSY in flat space, we may write the 1PI action as \cite{ArkaniHamed:1998kj,Giudice:1998xp}
\begin{align}
\mathcal{L} & \supset \frac{1}{4} \int d^4 \theta \,  \bRtilde(\Box) \bW^\alpha \left[ - \frac{1}{4} \D^2 \right] \Box^{-1} \bW_{\! \alpha} + \rm h.c., \label{eq:flatspace1PI}
\end{align}
or alternatively, remembering that $\frac{1}{16} \D^{\dagger 2} \D^2  = \Box$ when acting on chiral superfields, 
\begin{align}
\mathcal{L} & \supset \frac{1}{4} \int d^4 \theta \,  \bRtilde(\Box) \bW^\alpha \left[ - \frac{1}{4} \D^{\dagger 2} \right]^{-1} \bW_{\! \alpha} + \rm h.c. \label{eq:flatspace1PIb}
\end{align}
The superfield $\bRtilde$ (not to be confused with the chiral curvature superfield $\bR$) is the real vector superfield with the 1PI gauge coupling as its lowest component.  The dependence of $\bRtilde$ on $\Box$ encapsulates the running of the coupling with the momentum scale (selected by $\Box$, which should be thought of as acting only on the first $\bW^\alpha$).  A non-vanishing $\theta^2$ component for $\bRtilde$ yields a gaugino mass.  If $\bRtilde$ only has a lowest component, it then follows trivially that \Eq{eq:flatspace1PIb} is equivalent, after integrating over half of superspace, to the usual expression for the gauge kinetic lagrangian in chiral superspace (proportional to $\int d^2 \theta \, \bW^\alpha \bW_{\! \alpha}$).  

It is now a simple matter to generalize most of \Eq{eq:flatspace1PIb} to be SUGRA and super-Weyl covariant
\begin{align}
\mathcal{L} & \supset \frac{1}{4} \int d^4 \Theta \, \bE \, \bC^\dagger \bC \bRtilde(\superbox) \bWtilde^\alpha \P^{-1} \bWtilde_{\! \alpha} + \rm h.c., \label{eq:superweyl1PI}
\end{align}
where $\bWtilde_{\! \alpha} = \bC^{-\frac{3}{2}} \bW_{\! \alpha}$ has vanishing Weyl weight, and $\P$ is the super-Weyl covariant chiral projector given in \Eq{eq:Pdefinition}.  It can be easily verified that when $\bRtilde(\superbox) \bWtilde^\alpha$ is chiral, \Eq{eq:superweyl1PI} reduces to \Eq{eq:chiral1PI}, an integral of a local quantity over chiral superspace.

The only potentially ambiguous part of this equation is $\superbox$, the appropriately super-Weyl covariant version of $\Box$ acting on a super-Weyl inert superfield with an undotted spinor index.  If we only care about $\mathcal{O}(m_{3/2})$ effects such as gaugino masses, however, there are only two families of possible choices\footnote{For $\mathcal{O}(m_{3/2}^2)$ effects, such as non-local contributions to the self-energies of the particles in the vector multiplet (as considered in \Ref{ArkaniHamed:1998kj}), one would need to consider additional terms.  Such effects, the equivalents of the $\mathcal{S}$ and $\mathcal{T}$ of \Sec{subsec:supertrace} for vector multipets, are beyond the scope of this work. }
\begin{align}
\superbox \bU_{\! \alpha} & = \frac{1}{2} \bC^{\frac{1}{2}} \P \bC^{\dagger -1} \D_\alpha \frac{\D^\beta \bC^{\frac{3}{2}} \bU_\beta}{\bC^\dagger \bC} + \frac{1}{2} \bC^{\frac{1}{2}} \bC^{\dagger -1} \D_\alpha \frac{\D^\beta \bC^{\frac{3}{2}} \P \bU_\beta}{\bC^\dagger \bC} \nonumber \\
& \quad ~ + \frac{1}{4} \bC^{-\frac{1}{2}} \bC^{\dagger -1} \D^\dagger_{\alphadot} \D_\alpha \bC^{-1} \D^\beta \bC^{\frac{1}{2}} \D^{\dagger \alphadot} \bU_\beta \nonumber \\
& \quad ~ + a (\bP^\dagger) \P \bU_\alpha + \frac{1}{2} b \D_\alpha \frac{\D^\beta( \bC^{3/2} (\bP) \bU_\beta )}{\bC^\dagger \bC},
\end{align}
parameterized by arbitrary coefficients $a$ and $b$.\footnote{This is only gauge invariant for an abelian gauge theory; appropriate factors of $e^{\pm \bV}$ would need to be inserted for a non-abelian gauge theory.}  Note that the choice $a = 0$, $b = -1$ is especially convenient, as $\superbox \bU_\alpha$ is chiral for $\bU_\alpha$ chiral. This is precisely the choice used in \Eq{eq:superboxfirstdef}, and allows us to write the 1PI action as an integral over chiral superspace.  However, this choice is not necessary; regardless of the values of $a$ and $b$ chosen, a (more difficult) calculation shows that
\begin{align}
m_\lambda & = \frac{\beta_g}{g} m_{3/2}.
\end{align}

\bibliography{Scalars}
\bibliographystyle{JHEP}

\end{document}